\documentclass{aa}  

\usepackage{graphicx}
\usepackage{txfonts}
\usepackage{subcaption}         
\usepackage{lscape}             
\usepackage{placeins}           
                                
\usepackage{amsmath}	
\usepackage{bm}
\usepackage{amssymb}	
\usepackage{tabularx}
\usepackage{booktabs}
\usepackage{xcolor}
\usepackage[colorlinks=true,allcolors={blue}]{hyperref}
\DeclareMathOperator\erf{erf}

\newcommand{\dd}[1]{\,\text{d}#1}

\usepackage{xspace}
\newcommand{\HI}{\textsc{Hi}\xspace}

\begin{document}

\title{Painting a full radio sky}
\subtitle{Empirical mock catalogues with multiple source populations\\for future radio surveys}
%
%
%

\author{
T.~Ronconi\inst{1}\fnmsep\inst{2}\fnmsep\inst{3}\corrauth{tommaso.ronconi@inaf.it (INAF); tronconi@sissa.it (SISSA)}\email{to.ronconi@gmail.com}
\and
A.~Bonaldi\inst{4}
\and
M.~Spinelli\inst{5}
\and 
I.~Baronchelli\inst{1}
\and
M.~Behiri\inst{6}
\and
M.~Calabrese\inst{7}
\and
C.~Carbone\inst{8}
\and
M.~Giulietti\inst{1}
\and
A.~Lapi\inst{2,3,1}
\and
M.~Massardi\inst{1,2}
}
\institute{
INAF - Institute of Radio Astronomy (IRA), Via Gobetti 101, I-40129 Bologna, Italy 
\and
Scuola Internazionale Superiore di Studi Avanzati (SISSA), Via Bonomea 265, IT-34136, Trieste, Italy 
\and   
Institute for Fundamental Physics of the Universe (IFPU), Via Beirut 2, IT-34151, Trieste, Italy 
\and 
SKA Organization, Jodrell Bank, Lower Whitington, Macclesfield, SK11 9FT, UK 
\and
Observatoire de la Côte d’Azur (OCA), Boulevard de l'Observatoire, B.P. 4229, 06304 Nice cedex 04, France 
\and
INAF – Osservatorio di Astrofisica e Scienza dello Spazio (OAS), Via Gobetti 93/3, I-40129
Bologna, Italy 
\and
Astronomical Observatory of the Autonomous Region of Aosta Valley (OAVdA),  Loc. Lignan, 39, I-11020 Nus, Italy 
\and 
INAF - Institute of Space Astrophysics and Cosmic Physics (IASF), Via Corti 12, I-20133 Milano (MI), Italy 
}

\date{XXX; XXX}

\abstract{
Upcoming radio surveys will probe the sky with unprecedented depth and sky coverage, enabling a broad range of cosmological and astrophysical applications, as well as powerful synergies with experiments at other wavelengths. 
The preparation and scientific exploitation of these surveys require realistic mock catalogues that capture the complexity of the radio sky and the interplay of its emitting components.
In this work we present a modular and extensible algorithm for generating empirical simulations over the full radio sky, i.e. a solid angle of $4\pi$ steradians ($f_{\rm sky}=1$), down to redshift $z=5$, comprising both radio continuum and line emission. 
The framework combines a simulated dark-matter light-cone with empirically sampled galaxy populations and a probabilistic galaxy--halo assignment scheme, producing self-consistent mock catalogues including multiple radio populations on the same light-cone. 
We release two public catalogues: a shallow catalogue, fully constrained by existing observational data and limited to flux thresholds of $S_\text{1.4 GHz}^\text{lim} \sim 8\times10^{-5}\ \text{Jy}$ at $1.4\ \text{GHz}$ and $S_\text{21}^\text{lim} \sim 2\ \text{Jy}\cdot\text{Hz}$ for the \HI\ 21\,cm line; and a deep catalogue extending the calibrated empirical model to better sensitivities, broadly matching future SKAO surveys, with flux limits of $S_\text{1.4 GHz}^\text{lim} \sim 4\times10^{-5}\ \text{Jy}$ and $S_\text{21}^\text{lim} \sim 0.3\ \text{Jy}\cdot\text{Hz}$. 
The catalogues include radio continuum active galactic nuclei and star-forming galaxies, together with \HI-emitting galaxies, for a total of more than 260 million sources in the shallow catalogue and more than 1 billion in the deep catalogue.
We validate the simulations by analysing their statistical properties, distinguishing between quantities constrained by the empirical tuning and emergent relations from the combined modelling of baryonic populations and large-scale structure. 
The mocks reproduce the targeted clustering and population statistics while retaining minimal physical assumptions.
These simulations provide a flexible tool for survey forecasting, pipeline validation, and multi-tracer and multi-experiment synergy studies, and a foundation for extensions to alternative cosmologies, additional baryonic components, and multi-wavelength counterparts.
}
 

\maketitle
\nolinenumbers

\section{Introduction}

A new era for radio astronomy is approaching, driven by a suite of current and upcoming facilities that will dramatically expand both the depth and sky coverage of radio surveys, enabling radio cosmology as a fully developed field of study. 
This international effort will culminate in the SKA Observatory \citep[SKAO,][]{braun2019anticipatedperformancesquarekilometre}, 
which in its baseline configuration will consist of two complementary interferometric arrays designed for large-area and deep radio surveys: SKA-Low in Western Australia and SKA-Mid in South Africa. SKA-Low will comprise 131\,072 dipole antennas grouped into 512 stations operating over the $50$--$350\,\mathrm{MHz}$ range, while SKA-Mid will include 197 dishes (133 new SKA antennas plus the 64 MeerKAT dishes) covering frequencies from $\sim0.35$ to $\sim15\,\mathrm{GHz}$. 
With baselines extending to $\sim150\,\mathrm{km}$, wide instantaneous bandwidth, and sub-$\mu$Jy sensitivities achievable in deep integrations, the facility is designed to deliver wide, medium-deep, and ultra-deep surveys over cosmological volumes.
In anticipation of the SKAO, several precursor and pathfinder instruments are already producing large-area surveys and refining techniques critical for future science \citep[e.g.][]{precursors2024}, including the Australian SKA Pathfinder \citep[ASKAP][]{ASKAP2008,ASKAP2021}, the MeerKAT array in South Africa \citep{MeerKAT2016}, and the Murchison Widefield Array \citep[MWA][]{MWA2019}. 
Planned continuum and spectral-line surveys with these facilities, such as the Evolutionary Map of the Universe \citep[EMU][]{EMU2021} and the Rapid ASKAP Continuum Survey \citep[RACS,][]{RACS2024} on ASKAP and deep cosmological programmes with MeerKAT \citep[e.g., MIGHTEE,][]{MIGHTEEHI2021,MIGHTEECon2024} and LOFAR \citep[e.g.][]{LoLSS2023,LoTSS2023}, are detecting tens of millions of radio sources and are expected to characterise galaxy populations across cosmic time.
Finally, SKAO is expected to generate a coordinated set of wide, medium-deep, and deep surveys, targeting both continuum and spectral-line emission to address a wide range of cosmological and astrophysical science goals \citep{Redbook2020}.

Radio cosmology exploits these wide-area observations in the radio band to probe the structure and evolution of the Universe across cosmic time, using both spectral-line and continuum emission as tracers of large-scale structure.
Two complementary regimes can be distinguished: the high-redshift Epoch of Reionization (EoR), targeted primarily by low-frequency arrays aiming at the statistical detection of the redshifted 21-cm signal, and the post-reionization era, where mid-frequency instruments combine HI surveys and radio continuum observations to map the matter distribution and its evolution. 
Together, these approaches provide a unified framework investigating early astrophysical processes as well as late-time cosmology, including clustering and large-scale anisotropies.

In this context, realistic mock catalogues are an essential component for the preparation of surveys and to better interpret the coming data-sets. 
They are required to forecast survey performance, optimise observational strategies, validate analysis pipelines, and quantify statistical and systematic uncertainties. 
In the radio domain in particular, simulations must account for the coexistence of multiple astrophysical populations, complex selection functions, and strong redshift evolution.
As wide area surveys are expected to cover unprecedented sky fractions, preserving the large-scale clustering properties imprinted by the underlying matter distribution is also becoming a requirement. 

Several complementary approaches have been developed to model galaxy populations and their clustering. 
Hydrodynamical simulations provide a physically motivated description of baryonic processes, but are typically limited in volume and sky coverage \citep[e.g.][]{VillaescusaNavarro2018,Agertz2020,reina-campos2023,feldmann2023,schaye2025}.
Semi-analytic models offer greater flexibility, yet still rely on specific physical prescriptions that are difficult to validate across all observables \citep[e.g.][]{Croton2006,DeLucia2007,Henriques2015,Henriques2020,Fontanot2020,Fontanot2025,DeLucia2024,Parente2024,Yates2024}. 
Empirical simulations, on the other hand, efficiently reproduce selected observational statistics but often focus on single populations \citep[e.g.][]{Schreiber2017,Behroozi2019,Moster2018,Moster2021,Girelli2020} or lack a consistent embedding within a cosmological light-cone \citep[e.g.][]{Bonaldi2018,Bonaldi2023,Hartley2023}. 
As a result, no single existing approach simultaneously provides full-sky coverage, multiple radio populations, empirical flexibility, and modularity suitable for multiple survey-scale applications.

In this work, we present a modular pipeline for the construction of full-sky mock catalogues of observable radio objects, exploiting a strictly empirical approach.
Empirical models are anchored to observational constraints and make minimal assumptions about the physical mechanisms driving galaxy formation and evolution. 
While this limits their interpretability in terms of underlying physics, it ensures that the resulting simulations are not biased by incomplete or uncertain theoretical prescriptions. 
This makes empirical mocks particularly well suited for survey preparation, method validation, and cross-experiment synergy studies, where the goal is to reproduce observed phenomenology and its associated uncertainties rather than to provide ab-initio predictions.

Rather than constructing all components from scratch, we build upon and extend a set of well-established tools. 
The large-scale structure backbone is provided by a cosmological dark matter simulation through light-cones constructed from the high-resolution snapshots of the DEMNUni simulations \citep[Dark Energy and Massive Neutrino Universe,][]{Carbone2016,Hernandez-Molinero2024_a,Verza2024}, while mock galaxy populations are empirically sampled using the Tiered Radio Extragalactic Continuum Simulations framework \citep[T-RECS,][]{Bonaldi2018,Bonaldi2023}. 
These ingredients are combined through an extended version of the SCAMPy package \citep[Sub-halo Clustering and Abundance Matching in Python,][]{Ronconi2020}, which implements a probabilistic galaxy-halo assignment scheme to produce self-consistent mock sky catalogues. 
Radio continuum active galactic nuclei and star-forming galaxies on the one side, and \HI-emitting galaxies on the other, trace the underlying matter distribution in different ways and are selected by distinct observational criteria. 
Simulating these populations on the same light-cone enables joint analyses, cross-correlation studies, and multi-tracer applications that cannot be addressed using population-specific mocks in isolation.

This work also serves as a demonstration of the algorithm itself: its modularity offers in fact the possibility to extend the presented catalogues with further populations, coming, e.g., from observations at lower wavelengths, and thus fostering multi-instrument synergistic forecasts and studies.
The simulation pipeline is constructed such that its main components (the dark-matter backbone, the empirical galaxy sampler, and the galaxy-halo assignment scheme) can be modified or replaced independently. 
This allows the framework to be readily adapted to alternative N-body simulations, different empirical prescriptions, or additional galaxy populations as new observational constraints become available. 
As a result, the pipeline is not tied to a specific survey or modelling choice, but is intended as a flexible tool for a broad range of applications beyond radio-astronomy.

We construct two mock catalogues: a shallow catalogue, in which all model components are constrained by existing observational data, and a deep catalogue, which extends the calibrated model through controlled extrapolation to sensitivities broadly consistent with future SKAO survey capabilities. 
The resulting catalogues include radio continuum AGN and star-forming galaxies, as well as \HI line-emitting galaxies, all embedded within the same underlying dark-matter distribution.
The simulations presented here are designed for science preparation and methodological studies rather than for detailed physical modelling of galaxy formation.
The statistical properties of the catalogues are characterised and validated against available constraints, presenting a solid framework tuned on  the current empirical knowledge.

The paper is organised as follows. 
In Section~\ref{sec:ingredients} we describe the main ingredients of the model, including the dark-matter simulation, the empirical galaxy sampler, and the galaxy-halo assignment scheme. 
Section~\ref{sec:modelling} presents the simulation pipeline and the resulting mock catalogues. In Section~\ref{sec:properties} we analyse the statistical properties of the catalogues, distinguishing between quantities directly constrained by the empirical tuning and emergent relations arising from the combined modelling. We summarise our results and discuss future extensions in Section~\ref{sec:summary}.
\section{Ingredients of the model}\label{sec:ingredients}

In this Section, we describe the pipeline used to associate simulated radio sources to the halo/sub-halo hierarchy as derived from cosmological N-body simulations. 

In particular, on one side we have a simulated light-cone obtained by tiling different snapshots of a cosmological N-body simulation; we describe in detail the simulation used in Section~\ref{sec:demnuni}. 
On the other side, we have a catalogue of empirically simulated sources which mimic the statistical properties of observed sources; this is described in Section~\ref{sec:trecs}.
The aim of this work is to associate sources from the latter catalogue on top of the former light-cone based on observations of the clustering properties of the target populations of sources (i.e. continuum and HI radio galaxies).
The algorithm we have used to this purpose is described in Section~\ref{sec:scampy}.

\subsection{DEMNUNI Lightcones}\label{sec:demnuni}

In this work, we use the ``Dark Energy and Massive Neutrino Universe'' \citep[DEMNUni,][]{Carbone2016,Parimbelli2022} set of simulations.  These simulations have been produced using the tree particle mesh-smoothed particle hydrodynamics (TreePM-SPH) code Gadget-3 \citep{Springel2005}, and assume a Planck 2013 \citep{Planck2013} baseline flat $\Lambda$CDM cosmology with 15 different combinations of the total neutrino mass and the parameters characterising the dark energy equation of state. 
They are characterised by a volume of $(2 \, {\rm cGpc}/h)^3$ and $N_{\rm p}=2 \times 2048^3$ particles (with the $2$ factor representing neutrino particles when present), but here we consider only the massless neutrino flat $\Lambda$CDM case in its high-resolution run, with 64 times better mass resolution \citep[HR-DEMNUni, e.g.][]{Hernandez-Molinero2024_a, Verza2024}, characterised by a volume of $(500\ \text{cMpc}/h)^3$ and $2048^3$ CDM particles with no neutrino particles.

With snapshots of this simulation box we build a full-sky lightcone from redshift $z=0$ to redshift $z\approx8$, corresponding to a total volume of approximately $3\times10^3\ (\text{cGpc}/h)^3$ where we identify around 39 billion haloes and 44 billion sub-haloes by means of a Friends-of-Friends (FoF) algorithm and SUBFIND \citep{Springel2002}. The lightcone is constructed by embedding halo catalogues from the \texttt{DEMNUni} simulations into a full-sky, three-dimensional geometry centred on an observer, replicating the simulation volume to ensure continuous coverage along the line of sight. The volume is partitioned into concentric spherical shells of fixed comoving thickness, within which simulation outputs are coherently translated and rotated. This approach, originally developed for weak lensing applications~\citep{Carbone2008,Calabrese2015,Hilbert2020}, preserves the continuity of the gravitational potential across transverse directions.

The data we need from this light-cone are the position and masses of both the haloes and sub-haloes.
Nonetheless, the algorithm used to identify these structures from the simulation gives also access to a set of secondary properties such as circular velocity, concentration and environment.
Cosmological models assume that baryonic objects (i.e. galaxies) are hosted in sub-haloes which, conversely, are grouped into larger haloes (i.e. clusters or groups of galaxies).
Thus, by associating observables to N-body simulations, we also gain a statistically motivated description of the galaxy-halo connection.

\subsection{T-RECS}\label{sec:trecs}
The Tiered Radio Extragalactic Continuum simulation \citep[T-RECS,][]{Bonaldi2018,Bonaldi2023} produces catalogues of extragalactic sources with properties statistically consistent with current  observational constraints. Two models are available, briefly outlined in Sec.\ref{sec:continuum} and \ref{sec:hi}, dedicated to radio continuum and HI emission,  respectively. In terms of observational properties, users define their survey in terms of sky area and flux limit at a given frequency; outputs are produced for a list of user-defined frequency channels. 

\subsubsection{Continuum galaxies}\label{sec:continuum}
The T-RECS continuum model, described in details in \cite{Bonaldi2018}, covers the 150\,MHz and 20\,GHz frequency interval and includes Radio Loud Active Galactic Nuclei (RL AGN) and Star-forming galaxies (SFGs).  

RL AGN are based on the \cite{Bonato2017} evolutionary model, which classifies the AGN into steep-spectrum sources (SS-AGN), flat-spectrum radio quasars (FSRQs) and BL Lacs, with different evolutionary properties. Each sub-population is described by a redshift-dependent 1.4\,GHz luminosity function. Frequency behaviour of the sources is based on a power-law description, but in T-RECS the mean spectral index is constrained at different frequencies by the observed source number counts. 

SFGs are modelled as in \cite{mancuso2015} as late-type, spheroidals and lensed spheroidals. The radio emission is based on redshift-dependent star-formation rate (SFR) functions and a modelling of synchrotron, free-free and thermal dust emission as a function of SFR for each of the three sub-populations. The frequency dependence of the synchrotron emission, which is the dominant component, has been modified with respect to the original \cite{mancuso2015} model to ensure a good match with more recently observed number counts at different frequencies.

\subsubsection{HI galaxies}\label{sec:hi}
The T-RECS HI model is described in details in \cite{Bonaldi2023}.  The model rests on the $z=0$ HI mass function from \cite{jones2018}, combined with the \cite{duffy2012} relation between the HI mass and the total integrated HI flux of a source. 
The redshift evolution of the HI mass function is very poorly constrained at the moment. T-RECS models the evolution based on $z=0.3$ constraints from  
\cite{Bera2022} and \cite{Paul2023}. However, given the high uncertainties, the T-RECS HI model is limited to $z=0.5$, which limits the observable window to $\sim950\,\text{MHz}-1.4\,\text{GHz}$.
It is nonetheless possible, as new data from upcoming surveys arrive, to extend this validity window both in terms of frequency and redshift.

\subsection{The SCAMPy package}\label{sec:scampy}

SCAMPy \citep{Ronconi2020} is an extensible Python package designed to associate baryonic properties on top of the DM Halo/Sub-halo hierarchy from cosmological simulations.
The package implements the Sub-halo Clustering and Abundance Matching algorithm \citep[SCAM, e.g.][]{Guo2016}, which couples two widely used algorithms for the empirical definition of the galaxy-halo connection:
\begin{enumerate}
    \item an Halo Occupation Distribution (HOD, Section~\ref{sec:hod_model}) algorithm scans an input catalogue of DM haloes and sub-haloes and, among these, it chooses only those that will host an object of the target population.
    Whether a sub-halo will contain a baryonic object depends on the overall clustering properties of the target population.
    \item Sub-Halo Abundance Matching (SHAM, Section~\ref{sec:sham_model}) guarantees that each and every sub-halo in the simulation is associated with an object from the target population.
\end{enumerate}
In the next two Sections we describe in detail these two steps.

\subsubsection{HOD and Halo Model}\label{sec:hod_model}

In a hierarchical description of the Large Scale Structure (LSS) of the Universe, DM haloes contain substructures, dubbed sub-haloes.
While the former can be associated to clusters of galaxies and filaments, baryonic matter in-falls from these overdense environments towards the centres of the latter, while cooling down in the process and, eventually, forming galaxies.

This hierarchy of structures is found in DM-only cosmological N-body simulations \citep[e.g.][]{gadget, Springel2005}, and can be retrieved by means of clustering algorithms such as FoF and SUBFIND \citep{Springel2002} or ROCKSTAR \citep{rockstar2013}.

A HOD model predicts the average number of galaxies, $N_g(M_h, \boldsymbol{\theta})$, as a function of the host halo mass, $M_h$, and of a set of free-parameters, $\boldsymbol{\theta}$.
The function is usually expressed as the sum 
\begin{equation}
    \label{eq:Ngxy}
    N_g(M_h, \boldsymbol{\theta}) = N_\text{cen}(M_h, \boldsymbol{\theta}) + N_\text{sat}(M_h, \boldsymbol{\theta})
\end{equation}
of the average number of central galaxies $N_\text{cen}(M_h, \boldsymbol{\theta})$, taking on values from $0$ to $1$, and average number of satellites $N_\text{sat}(M_h, \boldsymbol{\theta})$.

A common and flexible parameterisation for these two functions is the standard 5-parameters HOD model \citep{Zheng2007,Zheng2009}, with the probability of having a central galaxy given by an activation function and the number distribution of satellite galaxies given by a power-law:
\begin{align}
    \label{eq:hod_cen}
    N_\text{cen}(M_h) &= \dfrac{1}{2}\left[1 + \erf\left(\dfrac{\log M - \log M_{\text{min}}}{\sigma_{\log{M_h}}}\right)\right]\\
    \label{eq:hod_sat}
    N_\text{sat}(M_h) &= \left(\dfrac{M_h - M_\text{cut}}{M_1}\right)^{\alpha_{\text{sat}}}
\end{align}
where $\erf(x)$ is the error function.
In this parametrisation, the halo occupation function of central galaxies is a transition from zero to one galaxy per halo, which is described by two parameters: $M_\text{min}$, the halo mass at which half of the haloes are populated by a central galaxy, and $\sigma_{\log M_h}$ , which characterises the smoothness of this transition. 
The satellites occupation function is instead modelled by a transition from zero galaxies to a power law with three parameters: $M_\text{cut}$,
the minimum halo mass for hosting satellites, $M_1 - M_\text{cut}$, the mass at which there is, on average, one satellite galaxy per host halo, and the power-law slope $\alpha$.

\begin{figure}
    \centering
    \includegraphics[width=0.48\textwidth]{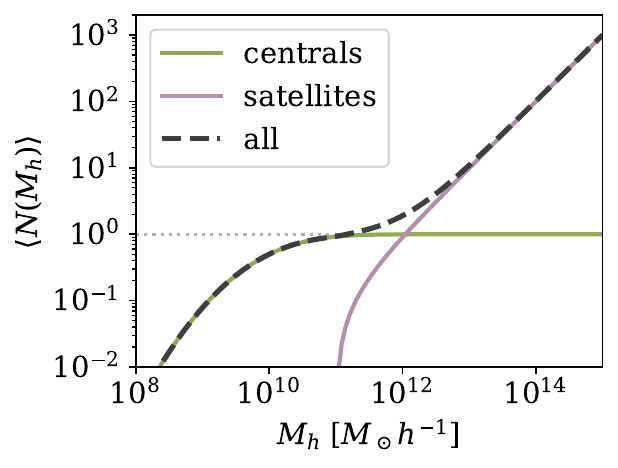}
    \caption{Average halo occupation distribution with the parameterisation defined in Eq.s~\eqref{eq:hod_cen} and \eqref{eq:hod_sat} and parameter values: $M_\text{min}=10^{10}~M_\odot h^{-1}$, $\sigma_{\log M} = 1$, $M_1 = 10^{10}~M_\odot h^{-1}$, $M_\text{cut}=10^{12}~M_\odot h^{-1}$ and $\alpha = 1$. The blue and green solid lines mark the contributes from central and satellite galaxies, respectively, while the dashed black line is the sum of the two. The dotted gray line highlights the threshold above which at least 1 galaxy is hosted, on average, by an halo of a given mass $M_h$.}
    \label{fig:hod_occupation}
\end{figure}
The overall average number of galaxies expressed by Eq.~\eqref{eq:Ngxy} is shown as a dashed black line in Figure~\ref{fig:hod_occupation}, while the contributes of central and satellite objects are marked by solid blue and green lines, respectively.
We have chosen an arbitrary set of parameters for plotting $N_\text{cen}$ and $N_\text{sat}$ in the Figure, with values $M_\text{min}=10^{10}~M_\odot h^{-1}$, $\sigma_{\log M} = 1$, $M_1 = 10^{10}~M_\odot h^{-1}$, $M_\text{cut}=10^{12}~M_\odot h^{-1}$ and $\alpha = 1$ (also reported in the Figure's caption).

Given a HOD model, it is possible to predict the clustering properties of a target population of galaxies using the halo model, which provides a theoretical approximated description of the non-linear evolution of matter perturbations.
The halo model stands on two assumptions:
\begin{enumerate}
    \item all matter in the Universe is partitioned within virialized haloes;
    \item for any matter tracer, the clustering properties at small scales are independent from the large scales.
\end{enumerate}
From this ansatz, the density field, in each position $\mathbf{x}$, is described as
the sum of the contributions of the density profile, $\mathbf{u}(\mathbf{x} - \mathbf{x}_i|m_i)$, of each halo with mass $m_i$, centred in $\mathbf{x}_i$: 
\begin{equation}
    \label{eq:density_hm}
    \rho(\mathbf{x}) = \sum_i m_i\, \mathbf{u}(\mathbf{x} - \mathbf{x}_i|m_i)\text{.}
\end{equation}
The very same formalism can be extended to any scalar field and can also describe cross-correlations between different tracers.

In our case, the scalar field we are interested in describing is the average number of galaxies, thus Eq.~\eqref{eq:density_hm} becomes
\begin{equation}
    \label{eq:ndensity_hm}
    N_g(V) = \sum_{M_{h,i} \in V} N_g(M_{h,i})\text{,}
\end{equation}
where $N_g(V)$ is the average number of galaxies within a volume $V$ and where $N_g(M_h)$ for the $i^\text{th}$ halo is parameterised as in Eq.s from \eqref{eq:Ngxy} to \eqref{eq:hod_sat}.
A more in detailed derivation of the 1- and 2-point statistics used in this work can be found in Appendix~\ref{apx:halo_model} and references therein. 

Before moving on, it is important to underline that both the assumptions mentioned above are strong and are known to severely limit the reliability of the clustering statistics predicted through the halo model \citep[e.g.][]{Acuto2021,Lacasa2022,Asgari2023}.
The halo model prediction is in fact flawed at the largest scales and inaccurate in the intermediate scales due both to the modelling limits and the strong assumptions; to correct for these effects several extensions have been proposed \citep[e.g.][]{Schmidt2016,Chen2020}.

Despite these known challenges, the halo model retains sufficient accuracy (and is routinely used) for calibrating the HOD parameters, enabling a reliable empirical description of halo occupation.

\subsubsection{SHAM}\label{sec:sham_model}

Abundance matching is a class of empirical techniques used to link galaxies to dark matter halos (or subhaloes). 
The method operates under the assumption that there is a monotonic relationship between a baryonic property (e.g., stellar mass or luminosity) and a halo property (e.g., halo mass or maximum circular velocity).

In practice, the cumulative distributions are computed for both the dark and baryonic properties. These distributions can be derived either from catalogues of objects (observed or simulated) or through analytical methods
\begin{equation}
    \label{eq:SHAM}
    \int_{O}^{+\infty} \dfrac{\dd{\Phi}(O)}{\dd{O}\dd{V}}\dd{V} \equiv \int_{M_h}^{+\infty} \dfrac{\dd{n}(M_h)}{\dd{M_h}\dd{V}} \dd{V}\,\text{,}
\end{equation}
where $\Phi(O)$ is the probability density function describing the average volume distribution of a baryonic property ($O$, e.g., the luminosity or the \HI mass of an object) while, on the right-hand side, the integrand contains the halo mass function.
By matching the integral of the two probability distributions, it is possible to obtain a bijective operative definition that associates each value in the dark sample with a corresponding value in the baryonic sample, $M_h\rightarrow O$.

In sub-halo abundance matching (SHAM), the relation is found for properties of the DM sub-halo instead of the parent halo.
Under the assumption that the HOD model is capable of predicting the total number of galaxies hosted by one DM halo, it is possible to match every single object from the sub-sample of sub-haloes obtained by filtering the original catalogue, with an  observable property which is either extracted from a catalogue (as those generated with T-RECS) or is obtained analytically by means of Eq.~\ref{eq:SHAM}.
\section{Pipeline and mock catalogues}\label{sec:modelling}

In this Section we describe the simulation pipeline and the modelling choices adopted in its construction. 
We outline the main components of the algorithm, including the dark-matter backbone, the empirical sampling of galaxy populations, and the galaxy-halo assignment scheme. 
We then summarise the key characteristics of the resulting mock catalogues and describe the observational datasets used to tune the empirical model parameters. 
Finally, we clarify the validity regime of the catalogues produced.

\subsection{Algorithm}\label{sec:algorithm}
The developed algorithm consists of three inter-playing steps which can be schematized with the following pipeline:
\begin{enumerate}
    \item Preparation of the ingredients:
    \begin{enumerate}
        \item Generate the halo and sub-halo catalogue of the reference DM light-cone (in this work we use the full-sky HR-DEMNUni light-cone described in Sec.~\ref{sec:demnuni}).
        \item Sample mock galaxy populations within the target sky area and redshift range (we use T-RECS to sample AGNs, SFGs and HIGs, as described in Sec.~\ref{sec:trecs}).
        \item Tune the HOD model on empirical data (we use the functions and models from the SCAMPy package described in Sec.~\ref{sec:scampy}). 
        Note that this step has to be performed for all the populations we want to model in our final mock catalogues, i.e., Radio-continuum sources and HI galaxies in this work (we give further details on the observational datasets selected for tuning in the following Section and in Appendix~\ref{apx:datasets}).
    \end{enumerate}
    These actions can be performed separately, though some mild degree of inter-dependency exists.
    Namely, in fitting the HOD parameters using the halo-model described in Sec.~\ref{sec:hod_model}, it has to be guaranteed that the cosmological parameters used are consistent with those of the DM simulation.
    Additionally, the resolution, size and redshift depth of the simulation also sets the survey area and depth of the T-RECS run. 
    \item For each population modelled, use the HOD to select haloes and sub-haloes from the DM simulation: this step produces a sub-catalogue of the original light-cone where all and only the sub-haloes that will host a mock galaxy of the target population are present.
    \item For each of the sub-catalogues perform the SHAM step of SCAMPy to associate a mock galaxy from the target mock sample with a DM sub-halo. 
    For this work, we match the cumulative sub-halo DM mass function of objects in the DEMNUNI light-cone with the T-RECS' 
    cumulative luminosity function at $1.4$ GHz, for the radio-continuum mock sources, and \HI mass function, for the \HI mock sources.
    For both the populations we introduce a stochastic scattering on the matching relation to account for second order effects.
    We apply the SHAM algorithm on a redshift-slice basis, in order to guarantee that the redshift distribution of the mock T-RECS sources is conserved.
\end{enumerate}
By applying the algorithm independently for each population considered on top of the same simulated DM light-cone, we automatically obtain the sky coordinates and the catalogue of counterparts of any simulated mock source in any other simulated catalogue.

We highlight that this algorithm is not limited to the ingredients we have selected in this work (i.e. DEMNUNI simulations and T-RECS samples) but it can also be applied to any other N-body simulation  \citep[such as the public FLAGSHIP simulation][]{EuclidFLAGSHIP2025} or approximated method for the production of DM halo/sub-halo hierarchies \citep[e.g. any method based on Lagrangian Perturbation Theory][]{Monaco2002, Monaco2013, Kitaura2013, Tassev2013, Tassev2015, Lepinzan2025}, and for any other empirical sampler of mock sources such as SPRITZ \citep{Bisigello_2021} or SEMPER \citep{Giulietti_2025}.    

The choice of the DM background simulation is driven by the scale of the astrophysical and cosmological processes we want to model.
The structural properties of the simulation determine both the resolution of the catalogues (i.e. the DM particle mass), which sets the lowest halo mass that can be reliably modelled (typically equivalent to the mass of some tens of DM particles), and the maximum reliable scale of the light-cone, which depends on the size of the original box used to tile up the light-cone.
While the former limit impacts on the maximum depth of the final catalogue, the latter limits the scale of the LSS cosmological probes for which the final catalogue can be relied on.
Note that this maximum scale has a fixed value in co-moving coordinates but it varies in spherical coordinates, as a result of the varying distance to the observer. 

\subsection{Catalogue products}\label{sec:catalogues}

\begin{figure*}
    \centering
    \includegraphics[width=0.98\textwidth]{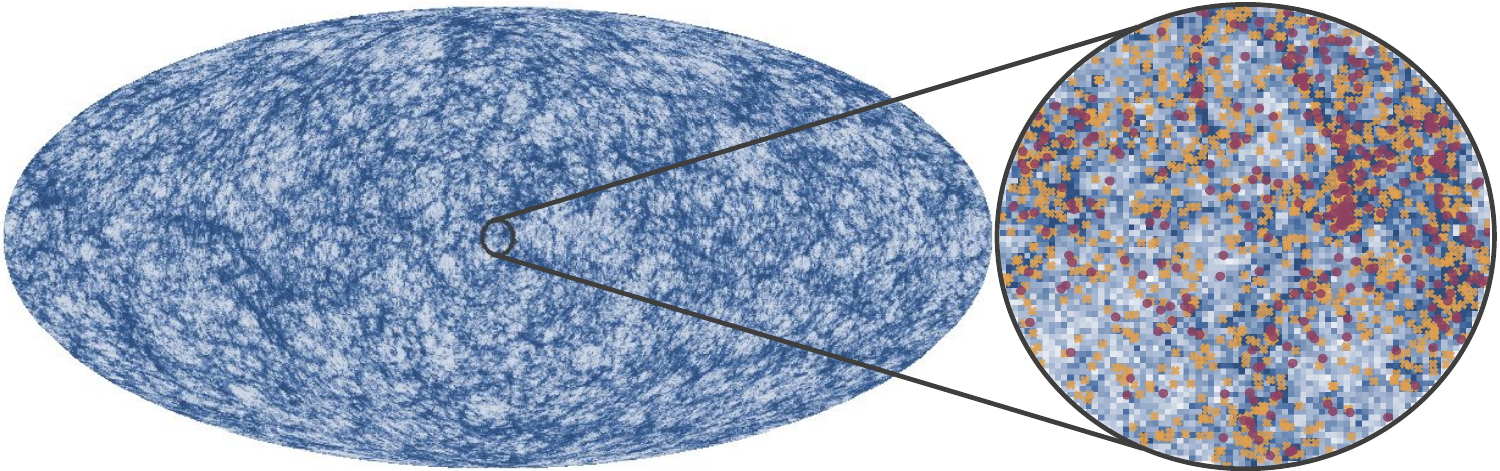}
    \caption{\textit{Left:} Projection of the whole light-cone in the redshift range $0 < z \leq 0.5$, the colour-scale shows the distribution of HI galaxies from lighter (lower number density) to darker (higher number density). \textit{Right:} FoV $= 12\ \deg$ zoom-in of the region marked by the black circle in the left-side projection showing the positions of AGNs (red markers) and SFGs (orange markers) distributed on top of the HIGs.}
    \label{fig:catalogue_zoomin}
\end{figure*}
Using the pipeline described in Sec.~\ref{sec:algorithm}, we can generate catalogues of different size, background cosmology, depth and astrophysical content.
In this work we release and present two different catalogues with different purpose:  
\begin{enumerate}
    \item a \textbf{shallow catalogue}, tuned to reproduce a realistic spatial distribution of sources by exploiting existing measurements of the two-point correlation function. 
    The parameters of the HOD model are therefore calibrated by maximizing a Gaussian likelihood that simultaneously accounts for both the one-point and two-point statistics, namely the observed number density of sources and their clustering signal:
    \begin{equation}
        \label{eq:chi2_1-2pt_scampy}
        \mathcal{L}(\boldsymbol{\theta}) = -\dfrac{1}{2}\left[\chi^2_{n_g}(\boldsymbol{\theta}) + \chi^2_{\xi_r}(\boldsymbol{\theta})\right],
    \end{equation}
    where $\boldsymbol{\theta}$ denotes the free parameters of the HOD model, $n_g$ is the spatial number density of sources, and $\xi_r$ represents the scale-dependent two-point correlation function in real space. 
    The observational datasets employed to constrain the HODs of the two main source populations in T-RECS are: the three-dimensional correlation function of \HI-selected galaxies from \citet{Martin2012}, and the angular correlation function of continuum-selected sources (SFGs and AGN) from \citet{Hale2018} (further details on the the dataset selection are presented in Appendix~\ref{apx:datasets}).
    The shallow catalogue is completely empirical, meaning that it does not contain extrapolations with respect to current knowledge from observed statistics. 
    \item A second, \textbf{deep catalogue} is generated by neglecting the clustering term in Eq.~\eqref{eq:chi2_1-2pt_scampy}, thereby omitting an explicit modelling of the two-point statistics. 
    In this case, the two independent HOD models are calibrated solely to reproduce the expected redshift evolution of the source number density, as predicted for a reference SKA radio survey.
    In particular, in continuum it is deep enough to both predict the source distribution for the Wide Band 1 Survey and the Medium-Deep Band 2 Survey as reported in \cite{CosmoSWG2020}; in \HI, we assumed a flux limit compatible with what expected from the Medium-Deep Band 2 Survey while accounting for the performance results of state-of-the-art \HI source detection software, as shown by the SKA Science Data Challenge 2 \citep{Hartley2023}.
\end{enumerate}

Both the catalogues contain the same populations of objects: radio continuum objects (CoG, hereafter), separated between AGN and SFG, and line-emitting neutral hydrogen galaxies (HIG) which mark the reservoir for the nuclear and star-formation activity of the former class of objects.
Additionally, in both catalogues each baryonic object also inherits all the host DM halo and sub-halo information of the original N-body simulation (e.g., mass, maximum circular velocity and coordinates).
Therefore, the difference between the two catalogues' \textit{flavours} is in the depth and validity regime of the astrophysical properties.
The shallow catalogue is strictly data-driven and valid within the calibration regime; the deep catalogue extends the model through controlled extrapolation to survey sensitivities relevant for the baseline SKA configuration \citep{braun2019anticipatedperformancesquarekilometre}. 
In Fig.~\ref{fig:catalogue_zoomin} we show the Mollweide projection of density perturbations in the neutral hydrogen density field of our shallow simulation integrated over the full light-cone where HIGs are defined (i.e. for redshift $z\in[0.0,0.5]$).  
In the right-hand side zoom-in we also show, on top of the \HI density field, the source distribution of CoG, divided in SFG and AGN.
In fact, by populating the same background DM simulation with sources of different type, both the shallow and deep catalogues automatically account for the existence of counterparts of objects of one population in the other.
This property is not explicitly modelled but is instead a result of having a shared ``background'' DM-only N-body simulation; we will comment the resulting statistics in Sec.~\ref{sec:cross_val}.

\begin{table*}
    \centering
    \begin{tabular}{l|rrr|r|rrr}
\toprule\midrule
  & \multicolumn{1}{c}{$N_\text{CoG}$} & \multicolumn{1}{c}{$N_\text{AGN}$} & \multicolumn{1}{c|}{$N_\text{SFG}$} & \multicolumn{1}{c|}{$N_\text{HIG}$} & \multicolumn{1}{c}{$N_\text{CoGxHIG}$} & \multicolumn{1}{c}{$N_\text{CoGxAGN}$} & \multicolumn{1}{c}{$N_\text{CoGxSFG}$}\\
  & \multicolumn{3}{c|}{$z \leq 5.0$} & \multicolumn{1}{c|}{$z\leq0.5$} & \multicolumn{3}{c}{$z \leq 0.5$}\\
\midrule
shallow & 217'143'484 & 30'553'719 & 186'589'765 &  51'509'884 &  7'084'155 & 1'067'733 & 6'016'422 \\
deep    & 365'527'186 & 40'607'134 & 324'920'052 & 719'501'958 & 32'531'743 & 2'156'916 & 30'374'827 \\
\bottomrule\bottomrule
\end{tabular}
    \caption{Summary of the catalogue sizes and flux limit. The first block of columns lists the exact numbers of objects in the continuum catalogue, the second for the HIG and the last block the sizes of the cross catalogues. Note that the CoG are divided between AGN and SFG both when listed alone and in the cross-catalogues column. The second line of the header lists the extension of the corresponding sub-catalogue in redshift and, in the last two columns, the physical units of the flux limit.}
    \label{tab:catalogues}
\end{table*}
The difference in total number of sources is also reported in Table~\ref{tab:catalogues}. 
While in continuum the deep catalogue has $50\%$ more sources than the shallow catalogue, while in \HI the deep catalogue has $\sim14$ times more objects than the shallow catalogue.
We also show in the last column the size of the cross catalogue, which is composed of all the sub-haloes hosting both a CoG and a HIG.
While in the shallow catalogue we find counterparts for $47\%$ of the CoG population and $14\%$ of the HIG population, in the deep catalogue this percentage grows up to $80\%$ of the CoG population and drops to less than $5\%$ of the HIG population\footnote{Note that the percentages mentioned have been computed accounting only for the $z\leq0.5$ sub-lightcone, where both the populations co-exist.}. 
This is due to the fact that the size of the continuum and \HI catalogues does not grow by the same factor going from shallow to deep.

The CoG in the shallow catalogue are complete down to flux $S_\text{1.4 GHz}^\text{lim} \sim 8\times10^{-5}\ \text{Jy}$ while the HIG down to line-flux $S_\text{21}^\text{lim} \sim 2\ \text{Jy}\cdot\text{Hz}$.
The deep catalogue is almost an order of magnitude deeper in both the populations with $S_\text{1.4 GHz}^
\text{lim} \sim 4\times10^{-5}\ \text{Jy}$ in continuum and $S_\text{21}^\text{lim} \sim 0.3\ \text{Jy}\cdot\text{Hz}$ in \HI.
We provide on the methology used to obtain these flux limits in Sec.~\ref{sec:derived}.
\section{Statistical properties}\label{sec:properties}

We have measured a set of relations between quantities available from the final mock catalogues in order to outline the statistical properties of the simulated samples.
We distinguish between quantities that are directly constrained by the empirical tuning procedure (presented in Subsection~\ref{sec:direct}) and quantities that emerge from the joint use of the DM simulation and the empirical galaxy models (presented in Subsection~\ref{sec:derived}).

\subsection{Direct results of the empirical model tuning}\label{sec:direct}

As described in Sec.~\ref{sec:algorithm}, the empirical model used to associate T-RECS mock-sources to DM sub-haloes accounts for two steps: in the first step we select DM sub-haloes onto which the mock-sources are mapped by the second step.
The parametrised HOD model is tuned differently for the shallow and deep catalogue: while the shallow catalogue is strictly phenomenological, aimed at reproducing the clustering properties measured by \cite{Hale2018} on the VLA-COSMOS field for CoG and by \cite{Martin2012} on the ALFALFA survey for the HIG, the deep catalogue is instead tuned on the abundance of sources at varying redshift expected from future SKAO surveys.

\begin{figure*}
    \centering
    \includegraphics[width=0.98\linewidth]{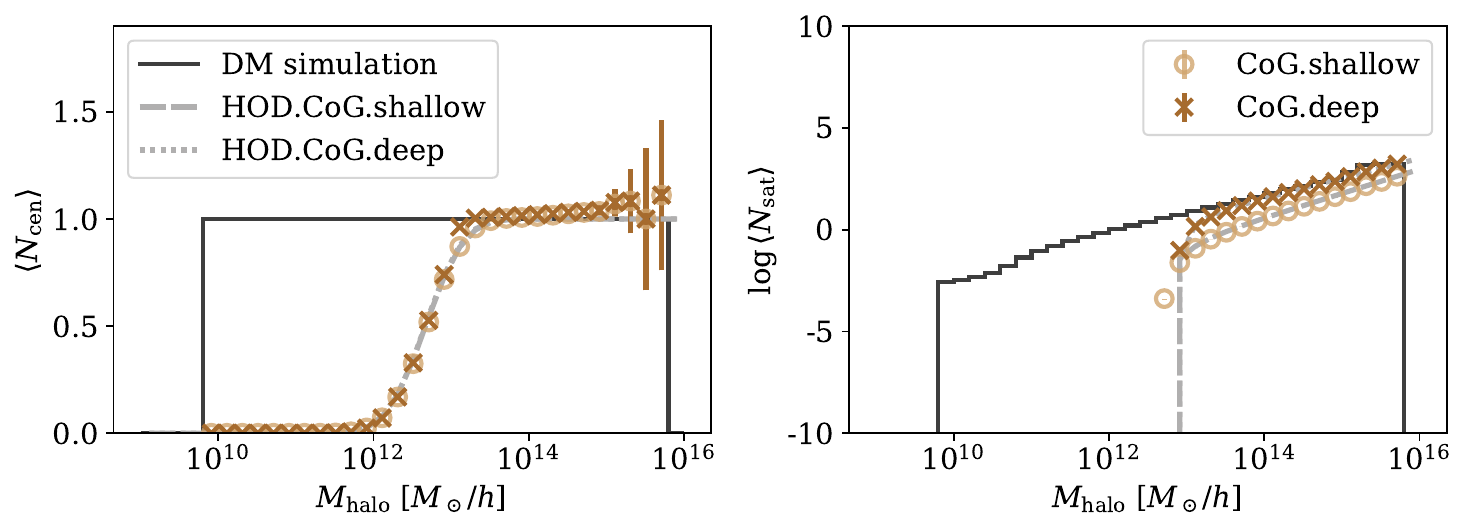}
    \caption{Average halo occupation distribution (HOD) separated between centrals (left panel) and satellites (right panel) as a function of the halo mass for CoG in the catalogues. The black solid line marks the distribution of sub-haloes within haloes in the original DM-only catalogue. The gray dashed and dotted lines mark the best-fitting HOD models obtained for the shallow and deep catalogue, respectively. Light orange circles and dark orange crosses mark the distribution of mock CoG in the final shallow and deep catalogue, respectively.}
    \label{fig:CoG_NcNs}
\end{figure*}

\begin{figure*}
    \centering
    \includegraphics[width=0.98\linewidth]{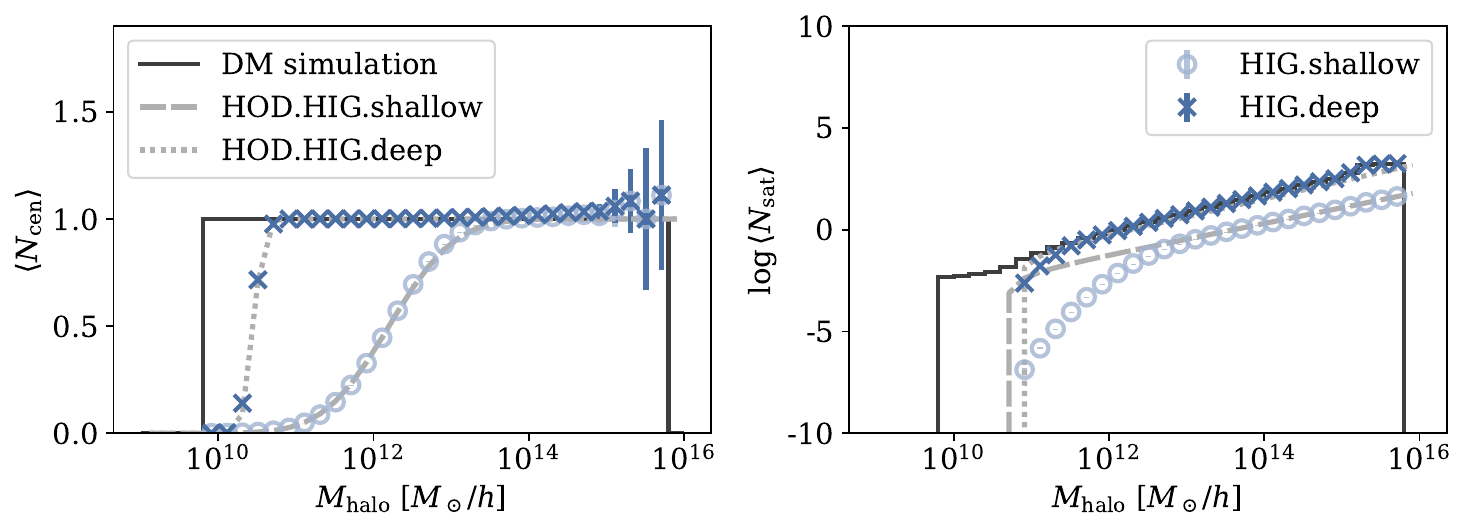}
    \caption{Same as Fig.~\ref{fig:CoG_NcNs} but for HIG. The measurement on the shallow and deep catalogue are marked by light blue circles and dark blue crosses, respectively.}
    \label{fig:HIG_NcNs}
\end{figure*}

\begin{table}
    \centering
    \begin{tabular}{lccccc}
\toprule\midrule
   & $M_\text{min}$ & $\sigma$ & $M_\text{0}$ & $M_\text{1}$ & $\alpha$\\
\midrule
\multicolumn{6}{c}{Continuum Galaxies}\\
\midrule
shallow & $4.69$ & $0.55$ & $5.55$ & $31.75$ & $1.19$\\
deep    & $4.61$ & $0.55$ & $8.05$ & $2.38$ & $0.97$\\
\midrule
\multicolumn{6}{c}{\HI Galaxies}\\
\midrule
shallow & $1.59$ & $0.94$ & $0.05$ & $40.49$ & $0.77$\\
deep    & $0.03$ & $0.15$ & $0.07$ & $0.30$ & $0.88$\\
\bottomrule\bottomrule
\end{tabular}
    \caption{Best fitting HOD model parameters in the 4 different cases considered, CoG and HIG in the shallow and deep catalogues. Note that the 5 free parameters of the model in Eq.s~\eqref{eq:hod_cen} and \eqref{eq:hod_sat} are fitted against different datasets and with different likelihoods in the 4 cases; further details are provided in Section~\ref{sec:catalogues} and in Appendix~\ref{apx:datasets}.different All the characteristic masses ($M_\text{min}$, $M_0$ and $M_1$) are in units of $[10^{12}M_\odot/h]$, the other parameters ($\sigma$ and $\alpha$) are dimensionless numbers.}
    \label{tab:HOD}
\end{table}
The HOD best-fitting parameters of CoG and HIG are reported in Table~\ref{tab:HOD} for both the catalogues (further details on the fitting procedure and priors are commented in Appendix~\ref{apx:datasets}).
Applying an HOD model to the N-body simulation results in a filtering of the full population of sub-haloes, where the total number of sub-structures present within a DM halo of given mass are sub-sampled to match an expected average abundance of hosts.
The result of this procedure is shown in Fig.~\ref{fig:CoG_NcNs} and Fig.~\ref{fig:HIG_NcNs}.

In particular, Fig.~\ref{fig:CoG_NcNs} shows the average number of central CoG (left panel) and satellite CoG (right panel) in the shallow catalogue (light orange empty circles and errors) and in the deep catalogue (dark orange crosses and errors).
By construction, the distributions measured in the 2 catalogues strictly follow the grey lines, which mark the best-fitting HOD models tuned for the CoG in the shallow and deep case (grey dashed line and grey dotted line, respectively).

As shown in the Figure, the distribution of sources in the 2 catalogues do not differ much between each other for what concerns central galaxies but show up to one order of magnitude more objects in the satellites average abundance.
Note that the black lines in both the plots mark the total distribution of sub-haloes in the underlying N-body simulation, thus imposing an upper limit to the maximum abundance of galaxies per given halo mass.
We have informed our fit of this limit; more details are reported in Appendix~\ref{apx:datasets}.

Fig.~\ref{fig:HIG_NcNs} shows the central and satellites distribution for the case of HIG, where the empty light blue circles and errors mark the distribution in the shallow catalogue, while the dark blue crosses and errors mark the distribution in the deep catalogue.
We stress again that, both HIG and CoG are built on top of the same DM simulation, therefore the solid black stairs in this Figure are precisely the same as in Fig.~\ref{fig:CoG_NcNs}.
In the \HI case, the two catalogues differ substantially both in the distribution of centrals and satellites.
This difference reflects the great improvement that SKAO observations will bring to the study of the neutral universe.
Also in this case we have imposed an upper limit set by the maximum number of available sub-haloes.
As is can be noticed by comparing the abundance of satellites measured on the shallow catalogue with respect to the corresponding model prediction, we tend to miss a part of the objects most likely as a result of sampling errors: the loss becomes relevant only for $\langle N_\text{sat}\rangle<1$, meaning that the average number of satellite galaxies expected within a given halo is smaller than one. 

\begin{figure*}
    \centering
    \includegraphics[width=0.49\textwidth]{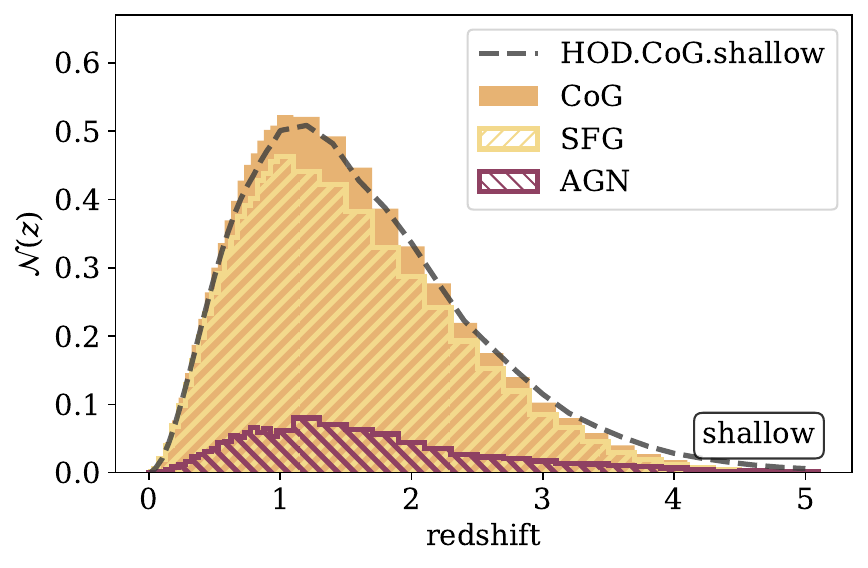}
    \includegraphics[width=0.49\textwidth]{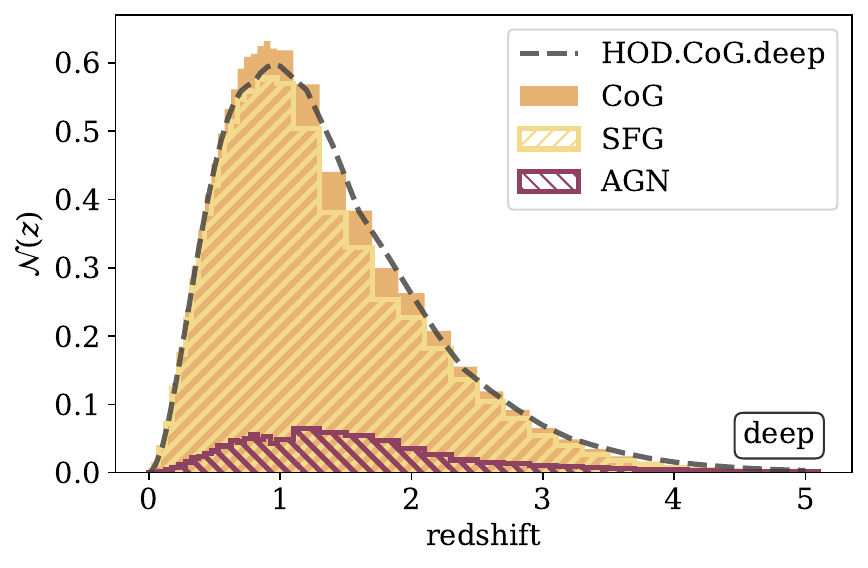}
    \caption{Redshift distribution of the CoG population in both the shallow simulation (left panel) and the deep simulation (right panel). Th normalized redshift distribution, $\mathcal{N}(z)$ of the whole population (orange) is divided into AGNs (hatched red) and SFGs (hatched yellow). The measurements on the simulated light-cones are compared against the prediction of the fitted halo model (dashed gray line).}
    \label{fig:CoG.zdist}
\end{figure*}
We show the normalized redshift distributions of CoG in the simulated light-cone in Fig.~\ref{fig:CoG.zdist}, measured in the shallow (right hand-side panel) and in the deep catalogue (left hand-side panel).
The measurement is reported as an orange histogram and compared to the prediction of the halo-model:
\begin{equation}
    \label{eq:Nz}
    \mathcal{N}(z) = \dfrac{\int_\Omega \dd{\Omega}\ n_g(z) \frac{\dd{V}}{\dd{z}\dd{\Omega}}}{\int_0^5\dd{z}\int_\Omega \dd{\Omega}\ n_g(z) \frac{\dd{V}}{\dd{z}\dd{\Omega}}}
\end{equation}
where $n_g(z)$ is the predicted density of sources at given redshift, Eq.~\eqref{eq:ngtot_hm}, and depends on the parametrised HOD model.
As expected the observed distribution perfectly matches the halo model prediction.
In both the panels we also separate the contribution of the two main populations of sources: AGN as a red hatched histogram and SFG as a yellow hatched histogram.
We notice that the distribution is slightly more peaked in the deep catalogue, implying that the additional source with respect to the shallow catalogue are mostly concentrated at lower redshift.

The time-complexity of direct estimators of the angular two-point correlation function (A2PCF) scales as $\propto N^2$, with $N$ the total number of sources.
As the number of sources projected on the full light-cone is extremely large, we have measured the A2PCF of CoG with a piece-wise approach: computing it separately for large-scales and small scales.
\begin{enumerate}
    \item The large-scale A2PCF has been measured by producing a HEALPix \citep{Gorski2005} map with $\text{N}_\text{side} = 2048$. 
    On this map we measure the angular power spectrum $C_\ell$ using the \texttt{healpy.sphtfunc.anafast} function of the \texttt{healpy} package \citep{zonca2025}.
    We then Fourier-transform the $C_\ell$ to obtain the A2PCF with
    \begin{equation}
        \label{eq:FT_Cell_a2pcf}
        \omega(\theta) = \sum_\ell \dfrac{2\ell + 1}{4 \pi} C_\ell P_\ell(\cos\theta)\ \text{,}
    \end{equation}
    where $P_\ell$ are the Legendre polynomials of order $\ell$.
    Errors on these estimator are computed as the square root of the variance
    \begin{equation}
        \label{eq:var_FT_Cell_a2pcf}
        \sigma^2(\theta) = \sum_\ell \left(\dfrac{2\ell+1}{4\pi}\right)^2 \text{Var}(C_\ell)P^2_\ell(\cos\theta)\,\text{,}
    \end{equation}
    where the variance on the $C_\ell$ is given by
    \begin{equation}
        \label{eq:var_Cell_a2pcf}
        \text{Var}(C_\ell) = \dfrac{2\,C_\ell^2}{(2\ell+1) f_\text{sky}}
    \end{equation}
    \item the small-scale clustering has been computed directly. We have divided the full light-cone into HEALPix pixels with $\text{N}_\text{side} = 4$ which corresponds to 192 regions of equal sky area. 
    On these regions, we apply the standard Landy-Szalay estimator 
    \begin{equation}
        \label{eq:landy-szalay}
        \omega(\theta) = \dfrac{DD(\theta)-2DR(\theta)+RR(\theta)}{RR(\theta)}\ \text{,}
    \end{equation}
    using the heaviside distance with radius $=1$ to compute the angular separations between tracers in both the data-catalogue and random-catalogue tiles.
    We then compute the mean and standard deviation of the measurement in each tile to get the final estimate and associated error for the small scales.
\end{enumerate}

\begin{figure*}
    \centering
    \includegraphics[width=0.49\textwidth]{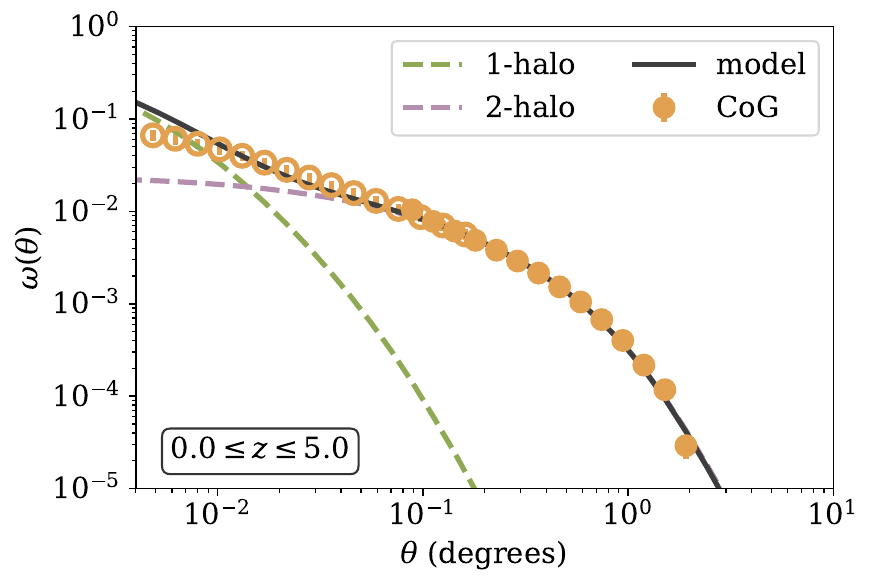}
    \includegraphics[width=0.49\textwidth]{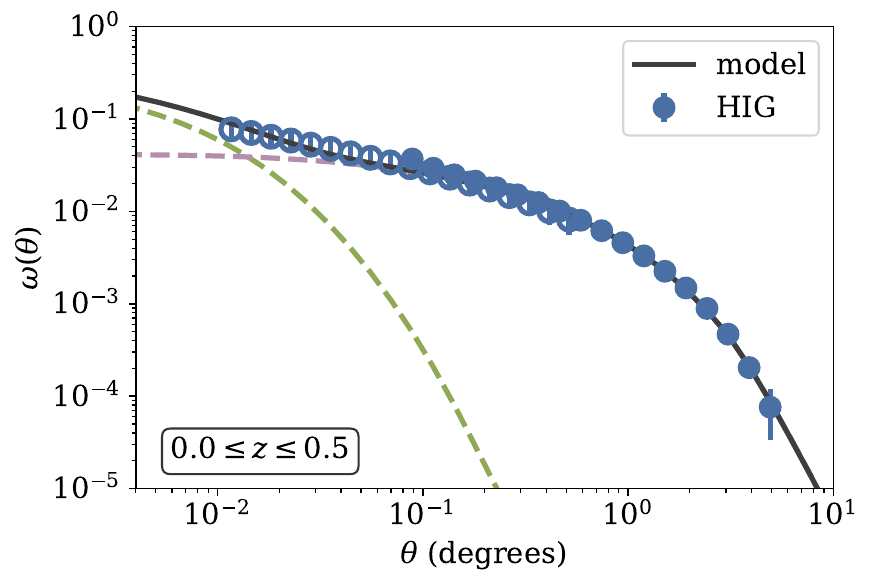}
    \caption{Angular 2 point correlation function measured on the shallow catalogue (markers), compared against the injected best-fitting halo model (solid line), divided into 1-halo (green dashed line) and 2-halo (magenta dashed line) contribution. The left panel shows the A2PCF of CoG and the right panel the A2PCF of HIG. In both panels, the lower left text-box shows the redshift limits of the considered population in the common light-cone. For both the dataset, we measured the A2PCF using different methods for large (filled markers and errors) and small scales (empty markers and errors).}
    \label{fig:a2pcf_agn+sfg}
\end{figure*}
Fig.~\ref{fig:a2pcf_agn+sfg} shows the A2PCF measured in the shallow catalogue compared against the fiducial halo-model prediction for both CoG (left panel) and HIG (right panel).
In both panels, the empty symbols mark the small-scale measurement performed with the Landy-Szalay estimator used for angular separation $\frac{1}{3}\sqrt{4\pi/N_\text{obj}}\leq\theta\leq0.5\ \text{deg}$, while the filled symbols are obtained by FT the angular power spectrum for separation $\theta \gtrsim 3\ \sqrt{4\pi/N_\text{pix}}$, with $N_\text{pix}$ the total number of pixel in the map used to estimate $C_\ell$.

Both in the CoG and in the HIG population extracted from the shallow catalogue, the measured A2PCF matches the halo-model prediction with the corresponding HOD parametrisation,
we can thus conclude that we are injecting the desired clustering properties in the catalogues, both in terms of abundance of sources for the deep as well as in terms of abundance of sources and 2-point correlation in the shallow catalogue.
The outcome of the SHAM step is instead commented in Appendix~\ref{apx:sham}.

\subsection{Cross-catalogue and emergent properties}\label{sec:derived}

The radio emission model is inherited from the T-RECS catalogues. 
As a consequence, all the scaling relations holding for the T-RECS sources are inherited by our light-cone, within the flux limits of the two catalogues produced.
Therefore here we limit our comment to the main statistics allowing to test the completeness of the catalogues and, in the following Section~\ref{sec:GXY2DM} and Section~\ref{sec:cross_val}, to the derived properties which result from the association of different populations to the same DM light-cone.
For the case of CoG, most of the statistics (including the A2PCF presented in the previous Section) are integrated along the redshift dimension.

\begin{figure*}
    \centering
    \includegraphics[width=0.98\textwidth]{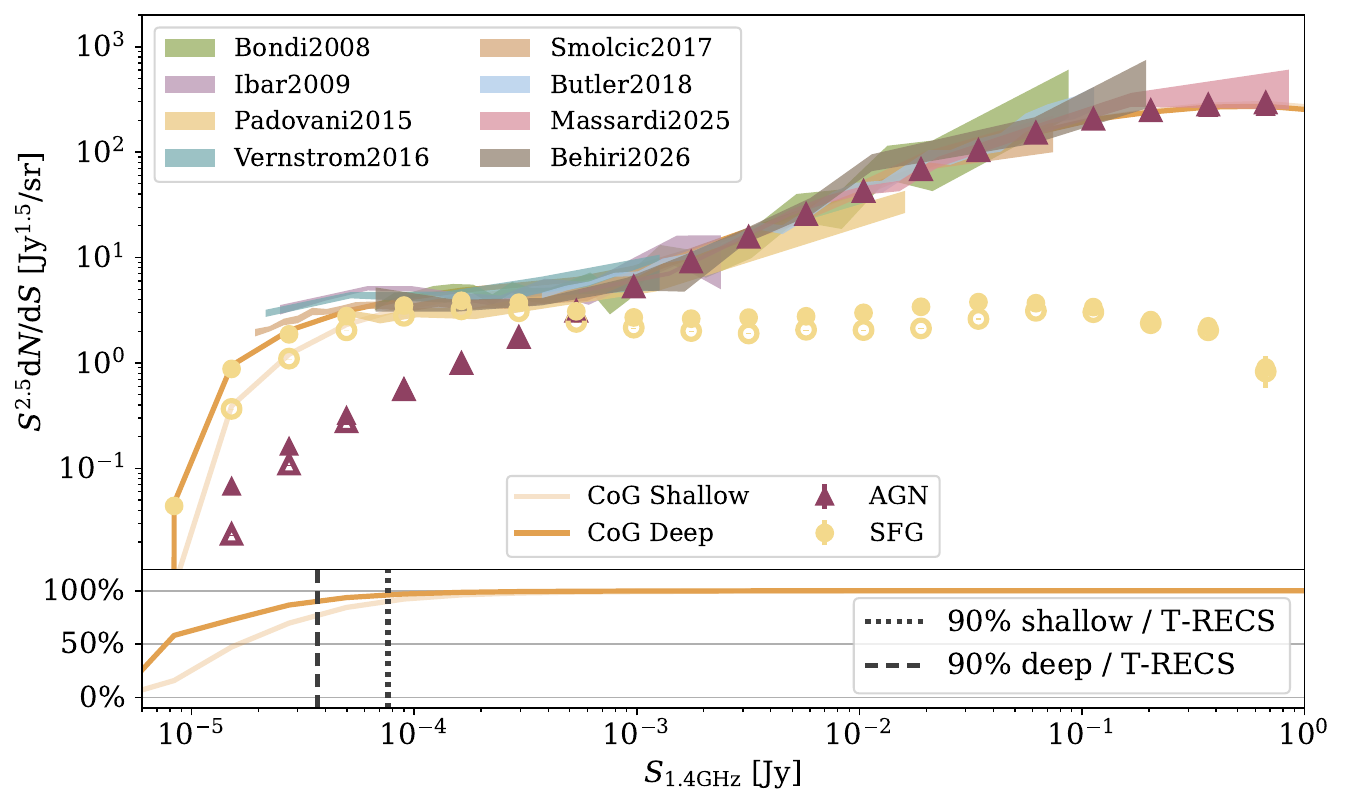}
    \caption{\textit{Upper panel}: Continuum galaxies differential number counts in source density at $1.4\ \text{GHz}$, measured on both the shallow and deep catalogues. The total distribution of sources in the simulated light-cones (solid lines, light orange for the shallow catalogue and dark orange for the deep catalogue) is divided in AGN contribution (red triangles) and SFG contribution (orange circles). Markers are empty for the shallow catalogue and filled for the deep catalogue. As a term of comparison, we also plot, as shaded regions, the 1-$\sigma$ confidence interval around available observational datasets from \citet{Bondi2008,Ibar2009,Padovani2015,Vernstrom2016,Smolcic2017a,Butler2018,Massardi2025,Behiri2026}.
    \textit{Lower panel}: Completeness of the CoG catalogue with respect to the cumulative number of T-RECS sources as a function of the emitted flux at $1.4$ GHz. We highlight the $90\%$ completeness value as a dotted black line for the shallow catalogue (light orange solid line) and as a dashed black line for the deep catalogue (dark orange solid line). 
    }
    \label{fig:CoG_number_counts}
\end{figure*}
In the upper panel of Fig.~\ref{fig:CoG_number_counts}, we compare the differential number counts of CoG to a set of observational constraints.
The highlight of this comparison is the possibility to visually extrapolate how well the simulated catalogues predict real observations.
In terms of upper limits, thanks to the wide simulated area, both the shallow and deep catalogue extend up to $S_\text{1.4 GHz}^\text{max} \approx 10^3\ mJy$, in agreement with the recent observations of \cite{Massardi2025}.


Producing a fair metric of source completeness, considering that all of this work is based on simulations, is not straightforward.
We provide indicative numbers by comparing the cumulative source count in both the shallow and  deep catalogues with the one measured in the T-RECS full catalogue used to generate them.
This measurement is shown in the lower panel of Fig.~\ref{fig:CoG_number_counts}, from which we see that the shallow catalogue reaches $90\%$ completeness at $S_\text{1.4 GHz}^\text{lim} \approx 0.08\ mJy$ (black dotted line) while the deep catalogue at $S_\text{1.4 GHz}^\text{lim} \approx 0.04\ mJy$ (black dashed line), thus improving in depth by a factor of 2 in flux limit and, in the deep catalogue, almost doubling the total number of sources.
The sources in our light-cone are therefore in reasonable agreement with both simulations and observation over more than 4 orders of magnitude in transmitted flux at $\nu = 1.4$ GHz.

\begin{figure*}
    \includegraphics[width=0.98\textwidth]{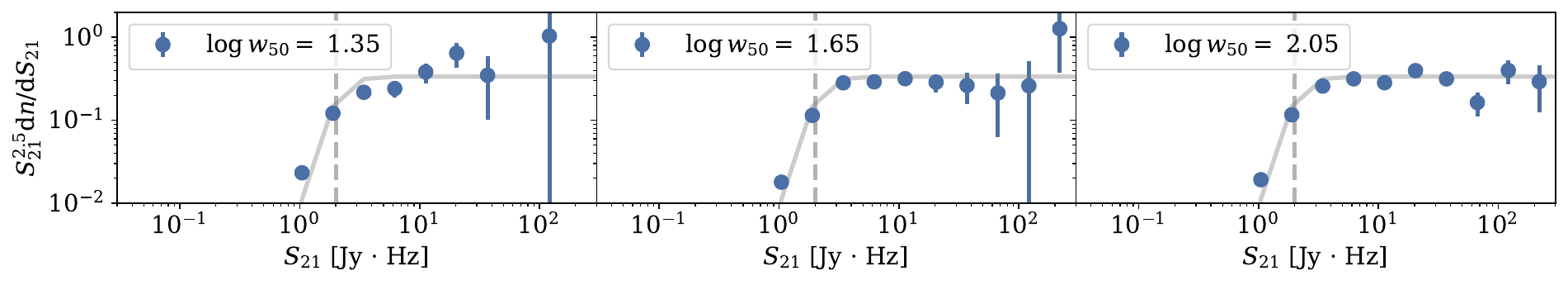}\\
    \includegraphics[width=0.98\textwidth]{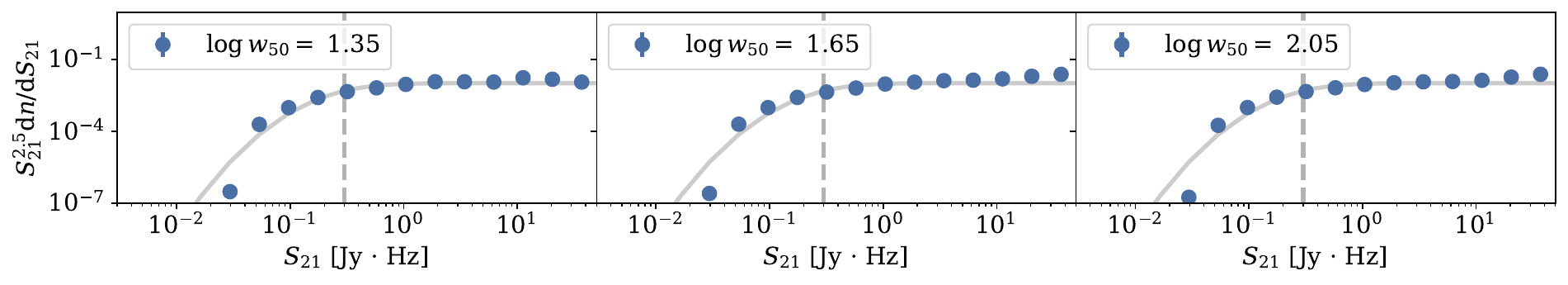}
    \caption{Completeness of the \HI population of sources in the shallow (upper panels) and deep (lower panels) in terms of the $S_\text{21}\text{-}S^{2.5}\dd{n}/\dd{S}$ distribution, shown for three representative values of the line-width $w_{50}$.}
    \label{fig:HIG_number_counts}
\end{figure*}
We also measure the differential number counts of HIG, for varying 21cm line flux in line-width bins $w_{50}$, to test the completeness of the two catalogues, as described in \cite{Haynes2011} and \cite{Ma2025}.
Since, at fixed $w_{50}$, the abundance of objects scales with $S_{21}^{2.5}$, the down-turn of the distributions shown in Fig.~\ref{fig:HIG_number_counts} marks the flux at which the catalogues' completeness falls below $100\%$.
In the Figure, the upper panels show the result measured (blue circles with errors) in the shallow catalogue, while the lower panels show the result for the deep catalogue, in the same $w_{50}$ bins (different columns).
We have fitted an error function (solid gray line) to the distribution, with shape
\begin{equation}
    \label{eq:HIGerfDist}
    S_{21}^{2.5}\dfrac{\dd{n}}{\dd{S_{21}}} \propto \dfrac{1}{2}\left[1+\dfrac{\log S_{21} - \log S_{21}^{50\%}}{\sigma_{\log S_{21}}}\right]
\end{equation}
with $\sigma_{\log S_{21}}=1.0$ and normalization tuned to match the plateau in the distribution.
The value of the free parameter $S_{21}^{50\%}$ marks the flux where the completeness falls below $50\%$ (marked).
In the shallow catalogue this value is $S_{21}^{50\%} = 2\ \text{Jy}\cdot\text{Hz}$, while in the deep catalogue it is almost an order of magnitude smaller: $S_{21}^{50\%} = 0.3\ \text{Jy}\cdot\text{Hz}$.

Besides the possibility of controlling the clustering properties of the resulting simulation of sources, another consequence of adopting our strategy for building simulated catalogues of sources, as described in Sec.~\ref{sec:algorithm}, is that cross-catalogues are generated without a-priori assumptions on the physics of baryons.
This means that the properties of the cross-catalogues are an indirect result of the abundance matching.
In the next two Sections we describe what we learn from analysing the cross-catalogues.
First, in Section~\ref{sec:GXY2DM} we disclose insights on the galaxy/halo connection which can be learnt from the catalogues. 
In the following Section~\ref{sec:cross_val} instead we focus on the counterpart cross-catalogue, which is obtained without any a-priori assumption besides and is a pure by-product of populating the same DM light-cone with multiple populations of sources.

We stress, however, that these cross-catalogues are not directly constrained by observational data, nor explicitly calibrated on the corresponding joint or conditional statistics. 
As a consequence, the inferred relations and trends should be interpreted with caution and not used to draw physical conclusions on the underlying baryonic processes. 
Rather, they reflect the internal consistency of the adopted modelling framework and the abundance-matching procedure. 
The present methodology is inherently flexible, and additional observational constraints on cross-population statistics can be naturally incorporated in future developments to improve the physical fidelity of these predictions.

\subsubsection{Galaxy/Halo connection}\label{sec:GXY2DM}

Even though the HOD step of the SCAM algorithm leverages on the total mass of the hosting halo it does not account for the baryonic content of such halo, but only on the number of sub-structures that are required to be visible to account for a given bias value.
On the other side instead, our implementation of the second SCAM step does not directly account for the total mass of the halo, nor for the sub-halo mass but is instead working only on a probabilistic level, thanks to the fact that, through the first SCAM step, all (and only) the sub-haloes that have to host an observable source have been isolated from the light-cone.
For this reason, all of the relations that link baryonic to dark quantities (besides those described in Sec.~\ref{sec:direct}) are emergent and not directly seeded into the catalogues.

\begin{figure*}
    \centering
    \includegraphics[width=0.98\textwidth]{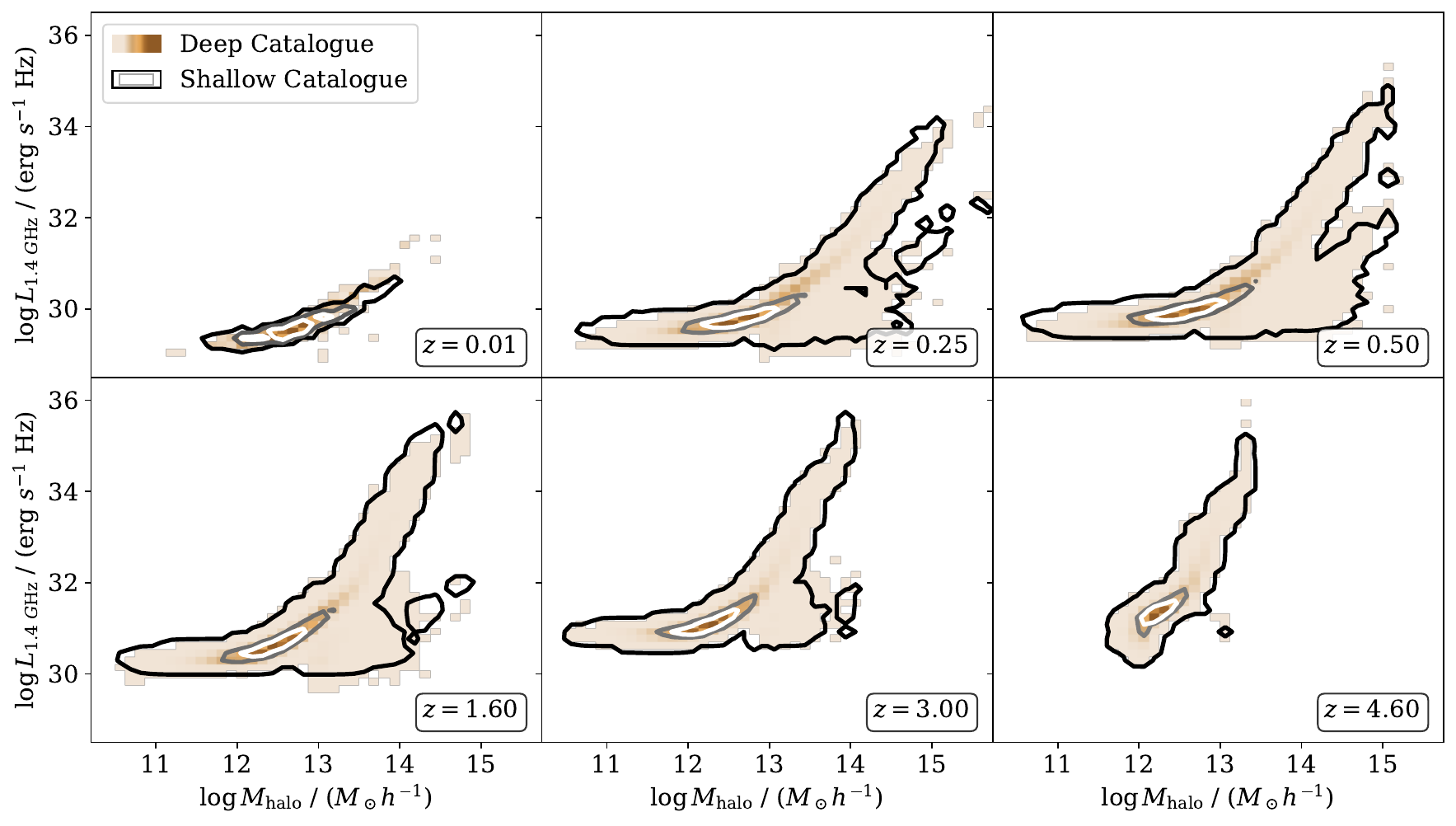}
    \caption{Integrated radio continuum luminosity at $1.4$ GHz against dark halo mass in the deep (shaded orange levels) and shallow (gray contours) catalogues. The results are plotted for 6 out of the 41 redshift slices available from $z_\text{min} = 0.01$ to $z_\text{max} = 5.0$. The central redshift of each slice is reported in the lower right box of each panel.}
    \label{fig:L1400vsMhalo}
\end{figure*}

In Fig.~\ref{fig:L1400vsMhalo} we show the total radio-continuum luminosity at $1.4$ GHz of DM haloes of a given size at different values of redshift (different panels).
In each panel we show both the scatter measured in the deep catalogue (shades of orange going from lower to higher percent of sources per bin) and in the shallow catalogue (grey contours highlighting $99.9\%$, $75\%$ and $50\%$ of all sources going from darker to lighter).
As shown in the Figure, there is almost no difference between the shallow and deep catalogue, which demonstrates that the contribution to the overall luminosity is dominated by the most luminous sources at all redshift, most likely AGNs.
The lower limit in luminosity, growing with increasing redshift is a direct result of the flux limit of the catalogue.

\begin{figure*}
    \centering
    \includegraphics[width=0.49\textwidth]{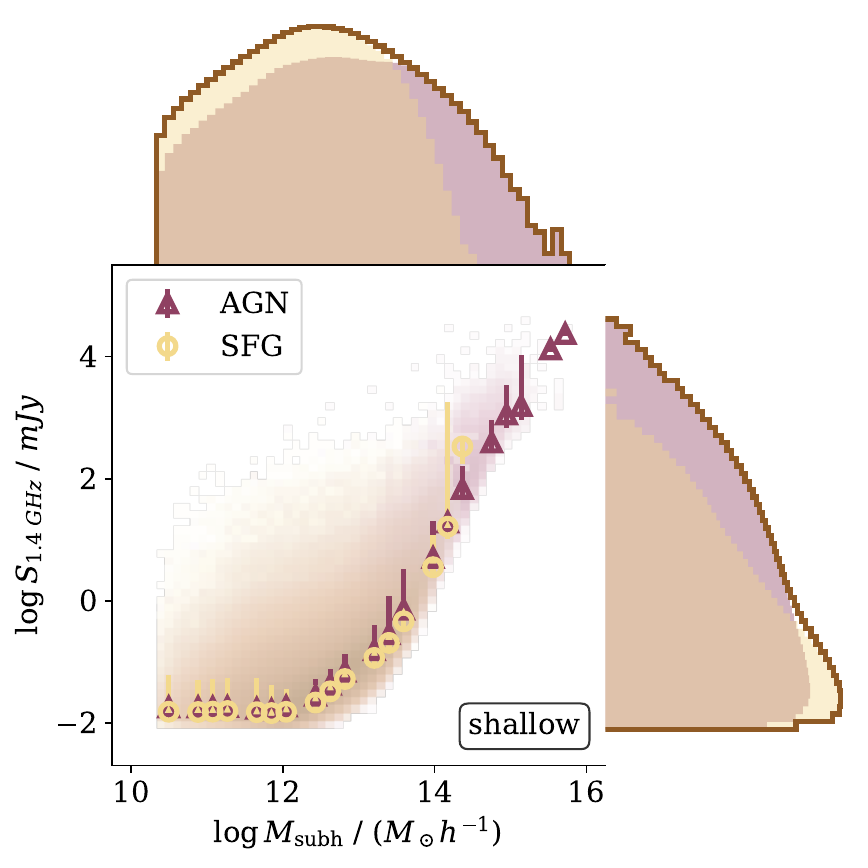}
    \includegraphics[width=0.49\textwidth]{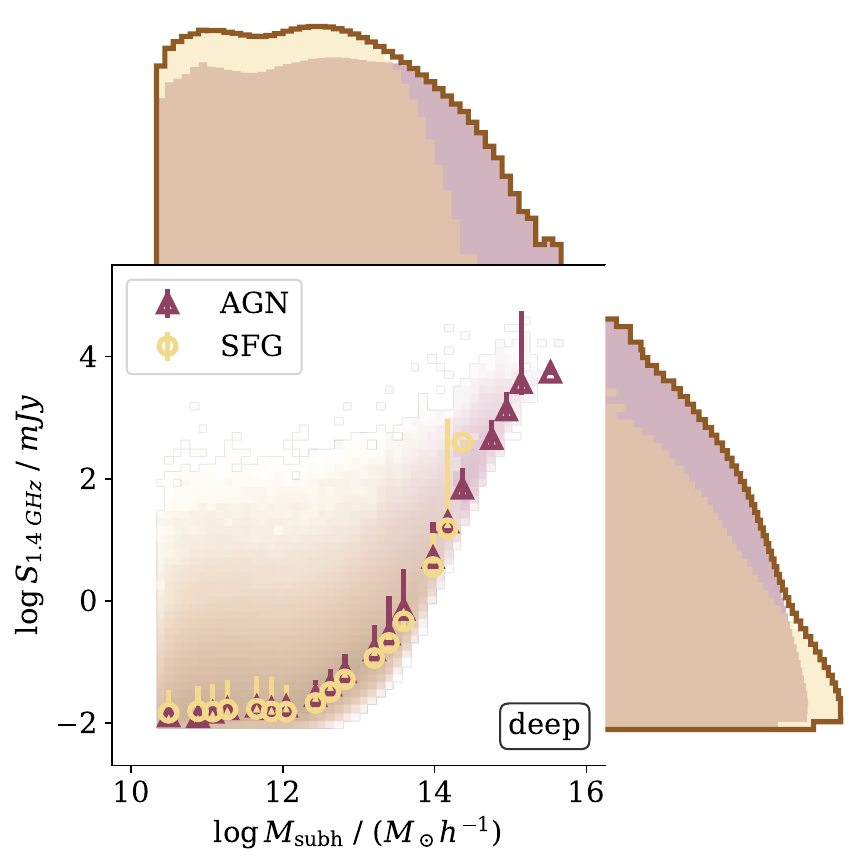}
    \caption{Flux at $1.4$ GHz integrated along the redshift dimension plotted against host sub-halo mass for the two population of sources in continuous catalogues. The two panels show the result for the shallow and deep catalogues (left and right, respectively), while the total logarithmic abundance of sources is histogrammed along the two dimension on the upper and right-hand sides of each panel. We separate between the distribution of AGN in red and SFG in yellow. With symbols and errors (empty red triangles for AGNs and empty yellow circles for SFGs) we mark the mean and standard deviation of the distribution, binned along the x-axis.}
    \label{fig:CoG_fluxVSMsubh}
\end{figure*}

The mass segregation between the two populations is more evident when plotted against the host dark mass, as done in Fig.~\ref{fig:CoG_fluxVSMsubh}, where the integrated flux to sub-halo mass is shown for the two different populations of sources. 
in the Figure, the shades of yellow and red show the percent distribution of sources in the continuum catalogue from the two different species: SFG and AGN, respectively.
We summarise the two distributions (in both the panels) with empty markers with error-bars, representing the mean and standard deviation along the y-dimension, averaged in bins along the x-dimension.
We plot results for both the shallow catalogue (right panel) and deep catalogue (left panel).

The two side histograms, on top and at the right of both the panels, show the projected distribution of the two populations along the sub-halo mass and flux dimensions, respectively.
As expected, we notice both a mass and flux segregation between SFGs and AGNs.
SFGs are typically more numerous at low values of mass and flux while the AGNs show a tendency on both properties to be shallower in their distribution at low values while reaching higher values of both mass and flux.
Actually, both the high mass and the high flux tails of the distribution are exclusively populated by this class of sources.

This plot highlights another perspective in the difference between shallow and deep catalogue, which is almost impossible to detect if it was not for the 1-dimensional distribution of object as a function of their host sub-halo mass which is shown on top of both panels.
The distribution in the shallow catalogue decreases below $\log M_\text{subh}/M_\odot h^{-1}\sim13$ while, in the deep catalogue shows a second peak at low dark masses.

\begin{figure*}
    \centering
    \includegraphics[width=0.98\textwidth]{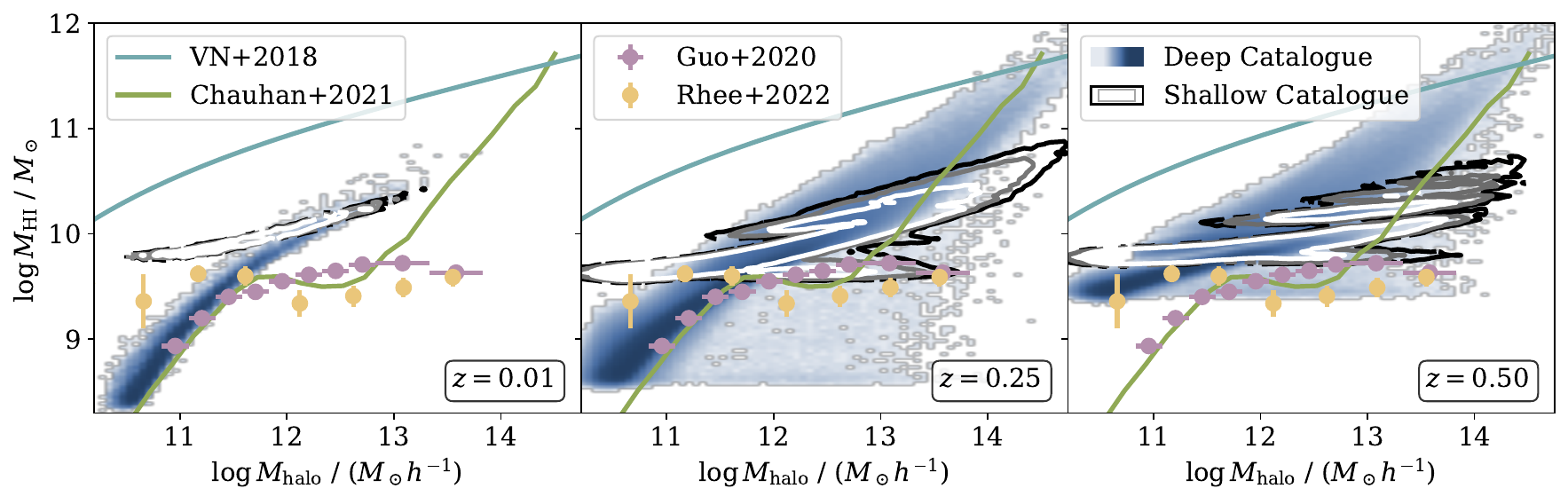}
    \caption{\HI to halo mass relation measured in the shallow catalogue (grey contours) and in the deep catalogue (blue shaded regions) in 3 redshift bins centred at the value shown in the lower left box of each panel. As a term of comparison, we also show results obtained from hydrodynamical simulations \protect{\citep[cyan and green solid line for][respectively]{VillaescusaNavarro2018, Chauhan2021}} and from measurements on real data \protect{\citep[coloured markers and error-bars][]{Guo2020,Rhee2023}}.}
    \label{fig:MHIvsMhalo}
\end{figure*}

Similarly to what obtained for the continuum catalogue, we can also inspect the \HI to halo mass relation, which is shown in Fig.~\ref{fig:MHIvsMhalo}.
The measurements performed in the shallow and deep catalogue are reported as grey contours (highlighting the $50\%$, $75\%$ and $99.9\%$ of sources from darker to lighter) and shaded blue region, respectively.
Each panel refers to a different redshift slice centred at the value reported in the lower left box.
In each panel we also over-plot results gathered from literature, either on simulations (solid lines) or on actual observational data-sets (markers and errors).
Our catalogues are compatible with all the other measurements reported in the Figure, apart for the \cite{VillaescusaNavarro2018} model.
We also note that the deep catalogue is more representative of observations, especially at the lower redshift and in the lower \HI mass region of the plot.
While the \cite{Chauhan2021} relation is obtained through hydrodynamical modelling, both \cite{Guo2020} and \cite{Rhee2023} produce their measurement through stacking of the \HI $21\ cm$ line around redshift $z\sim0$.
The good agreement of our measurements with these datasets, especially at the lower redshifts, validate further the empirical reliability of our catalogues.

Along with the several quantities provided by T-RECS, the mock HIGs produced by the sampler include information on their maximum circular velocity \citep[based on the empirical relation of][]{Katz2018} and on the resulting emission line width.
Given that sub-haloes in N-body simulations come with information on the velocity of the test particles used to model them, in our catalogues the maximum circular velocity is also available from the backbone N-body simulation.
\begin{figure}
    \centering
    \includegraphics[width=0.98\linewidth]{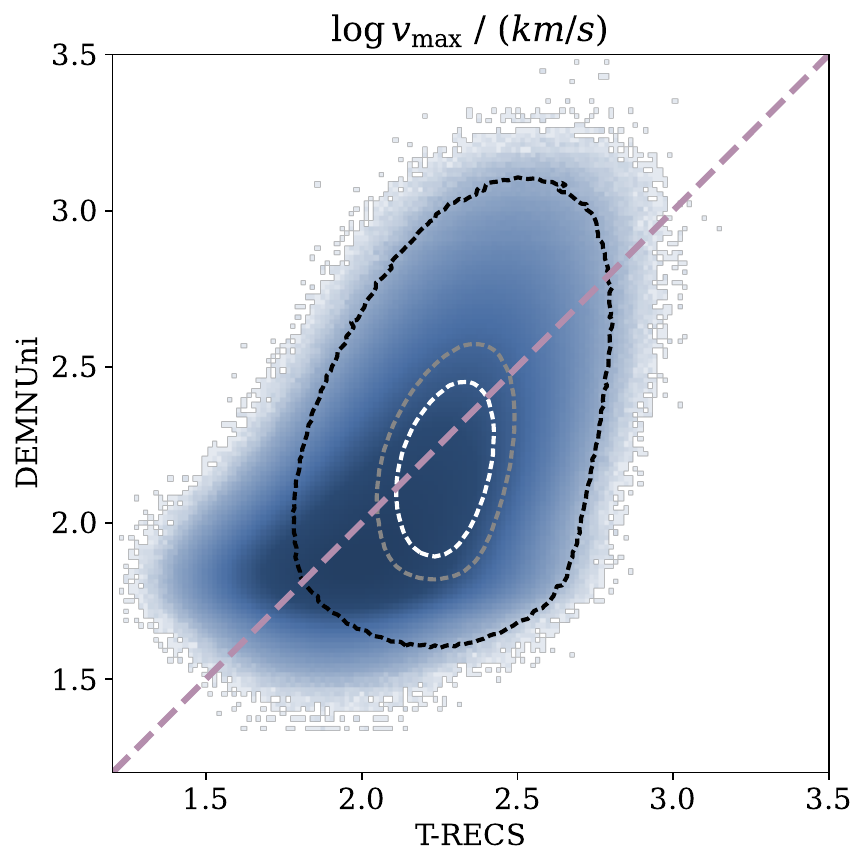}
    \caption{Relation between the maximum circular velocity predicted by the T-RECS library and the one measured on the N-Body sub-haloes in the DEMNUni light-cone, for all the HIG in the catalogue. The gray contours and the coloured region show the log-scaled percent abundance of sources in the shallow and deep catalogues, respectively, while the dashed line marks the $y = x$ relation. The axes are in $\log\ km/s$ units.}
    \label{fig:DEMNUNIvsTRECS_Vmax}
\end{figure}
In Fig.~\ref{fig:DEMNUNIvsTRECS_Vmax} we plot the relation between the value of $v_\text{max}$ as predicted by the two modelling strategies (gray contours and coloured shaded region, depending whether measured in the shallow or in the deep catalogue, respectively).
By comparing with the $y = x$ relation, it is evident that the two are compatible, even though the distribution shows a large scatter.
The advantage of using the information on $v_\text{max}$ from the DM sub-haloes instead of the one provided by T-RECS is that the latter only predicts the quantity for HIG, while the former is available for all the mock sources, given that our algorithm has linked each observable mock source to a specific sub-halo in the light-cone.
At the same time, the agreement between the two predictions shown in Fig.~\ref{fig:DEMNUNIvsTRECS_Vmax} guarantees that this is a safe assumption. 

Since the information on the inclination angle of HIGs with respect to the line of sight is also available from the T-RECS catalogue, we use Eq.~(7) of \cite{Bonaldi2023}
\begin{equation}
    \label{eq:w50}
    w_\text{50} = 2 \sin(i)\,v_\text{max}
\end{equation}
to compute the line broadening due to rotation of the HIG as resulting from the circular velocity given by the DM simulation. 
We have included this additional estimate in the final catalogues.

\begin{figure*}
    \centering
    \includegraphics[width=0.98\textwidth]{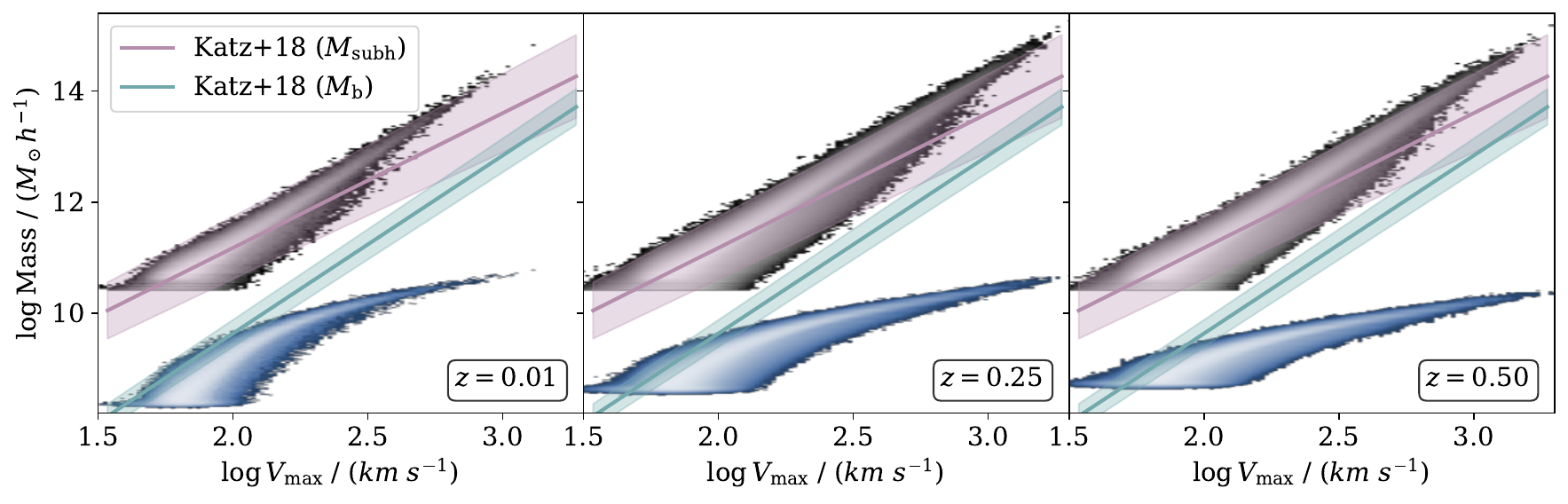}
    \caption{
    Dark (grey shaded region) and baryonic (blue shaded region) mass relation with respect to the $v_\text{max}$ of sub-haloes of the DEMNUni N-body simulation.
    The $M_\text{subh}\text{-}v_\text{max}$ (magenta solid line and $1\text{-}\sigma$ confidence) and $M_\text{b}\text{-}v_\text{max}$ (green solid line and $1\text{-}\sigma$ confidence) from \protect\cite{Katz2018} are also reported for comparison. 
    The lower right text box in each panel shows the central redshift of the light-cone slice considered.}
    \label{fig:massVSvmax}    
\end{figure*}
We provide a final test on the new association of quantities obtained by linking the observables to the underlying DM sub-haloes by plotting, in Fig.~\ref{fig:massVSvmax}, the sub-halo mass and \HI mass against the $v_\text{max}$ value obtained from the N-body simulation as a grey shaded and blue shaded region, respectively, for all the redshift slices in the catalogue (redshift centres are reported in the lower left text box).
The distributions are compared with the prediction from \cite{Katz2018} on which the T-RECS model is tuned.
The purple line with associated $1\text{-}\sigma$ confidence marks the $M_\text{subh}\text{-}v_\text{max}$ relation which is almost perfectly matched by the sub-halo properties, at all the redshift available. 
The higher mass end of the relation (i.e. $M_\text{subh}\gtrsim10^{13} M_\odot$), even though following the general trend and being within the $1\sigma$ confidence region of the empirical relation (red shaded region), shows a concentration of objects above the mean (solid magenta line).
This is not concerning as in the original work, the relation is tuned on haloes with mass lower than this limit.

On the other hand, the $M_\text{b}\text{-}v_\text{max}$ between the baryonic mass and maximum circular velocity measured from \cite{Katz2018} (green solid line and associated $1\text{-}\sigma$ confidence) is only representative of the lower mass hand of the \HI catalogue.
This is not surprising as in most of the galaxies \HI is not sufficient to account for all the baryonic mass, which is instead distributed between stars, molecular and ionized regions.
The measurement in our catalogues suggests that \HI alone is not simply a biased proxy of the whole baryonic matter contained in sub-haloes, but there is instead a dependence on the $v_\text{max}$ (or conversely, the sub-halo mass) of the biasing factor.

\subsubsection{Baryonic relations}\label{sec:cross_val}

Since in our empirical model we do not force any relation between the different populations modelled (continuum and line-emitters), it is interesting to inspect how the observable quantities are correlated in our emergent cross-catalogue.

The \HI mass is generally considered to be a good proxy of star formation \citep[e.g.][]{Michalowski2015,Naluminsa2021}.
In the original T-RECS sampler, where a specific recipe for cross-matching between catalogues is implemented, the latter relation is used to predict from the continuum catalogue the expected $M_\text{HI}$, which is then used to find in the \HI catalogue the mock source with the value of this property which is closer (within some error) to this prediction.

\begin{figure*}
    \centering
    \includegraphics[width=0.95\textwidth]{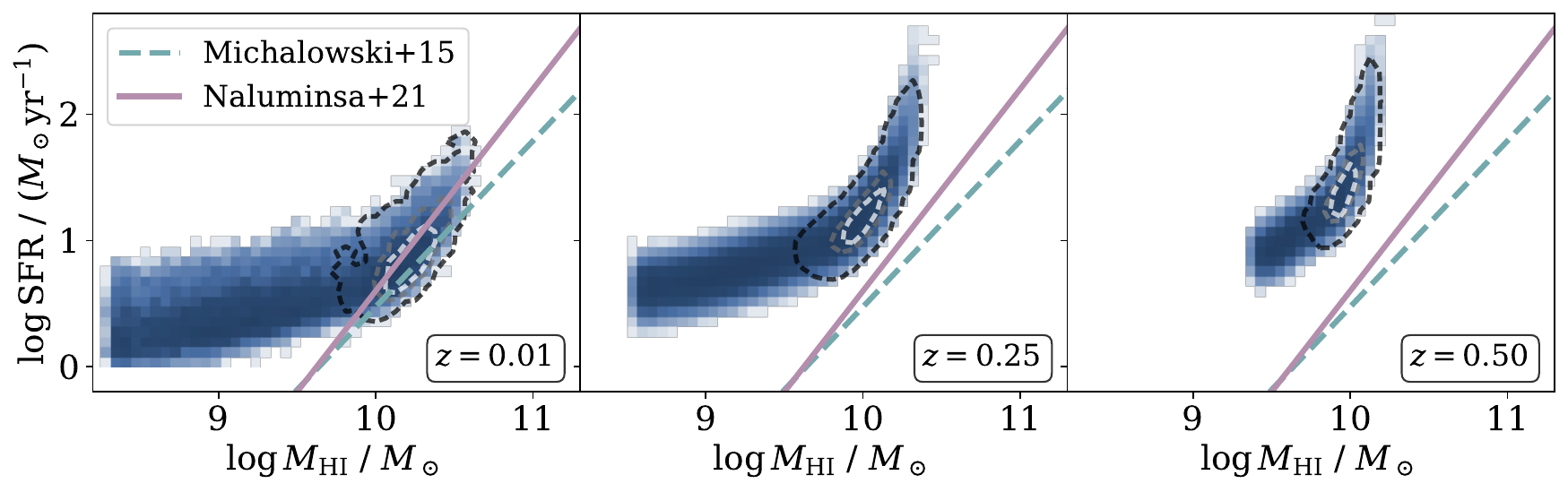}
    \caption{Distribution in each redshift slice or the relation between the neutral hydrogen mass and the star formation rate of each object in the cross catalogue. As a term of comparison, we also show 2 results fitted from literature \protect{\citep[i.e.][as a cyan dashed line and a pink solid line, respectively]{Michalowski2015,Naluminsa2021}}.}
    \label{fig:MHIvsSFR}
\end{figure*}
We do not impose any of this in our catalogues but, in Fig.~\ref{fig:MHIvsSFR}, we plot the emergent relation for both the shallow (gray contours) and deep (shaded blue regions) catalogues.
The different panels show the distribution measured at different redshift values, as reported in the lower right text-box.
In each panel we also plot the observed empirical relations from both \citet[][dashed cyan line, M15 hereafter]{Michalowski2015} and \citet[][solid magenta line, N21 hereafter]{Naluminsa2021}.
At redshift $z\lesssim0.1$, both the shallow catalogue and the higher \HI mass end of the deep catalogue are in a good agreement with the empirical relations while they move farther from the lines as we move to the higher redshift slices.
The M15 relation is obtained from a sample of HIG extending up to redshift $z<0.12$, while N21 encompasses a light-cone with depth $r\approx72\ Mpc/h$ (which, in the cosmology adopted here, corresponds to $z\lesssim0.024$).
It is therefore expected not to match the empirical relations at higher values of redshift.
This behaviour is possibly due to the strong evolution of the \HI mass function model adopted in the T-RECS sampler, which is extrapolated from the $z=0$ model of \cite{jones2018} up to redshift $z=0.5$, leveraging on early estimates \citep{Bera2022,Paul2023} of the mass distribution of \HI sources at redshift higher than $z\gtrsim0.3$. 

It is instead less expected the change in slope observed at lower values of $M_\text{HI}$, which contains the bulk of our population of cross-matched mock sources.
From comparing with the mass limits of the sample used to fit the reported relations, especially considering the N21 relation.
Even though the sample encompasses sources with mass down to $\log M_\text{HI}/M_\odot \approx 7.5$, the bulk of the population is concentrated at relatively higher \HI masses (i.e. $\log M_\text{HI}/M_\odot\gtrsim 8.5\text{-}9$).  
Anyway, this does not appear to be enough to motivate the tilt of distribution we observe in the deep catalogue. 
We plan to investigate further this behaviour in follow up work on this topic.

\section{Summary}\label{sec:summary}

In this work we have presented a modular, empirical pipeline for generating full-sky mock catalogues of the radio sky, designed to support science preparation and methodological studies for current and forthcoming radio surveys. 
The framework combines a cosmological dark-matter (DM) light-cone with empirically sampled galaxy populations and a probabilistic galaxy–halo assignment scheme, producing self-consistent simulations of radio continuum active galactic nuclei (AGN) and star-forming galaxies (SFG), as well as 
neutral hydrogen (\HI) galaxies, embedded within the same large-scale structure realisation.

We constructed two mock catalogues corresponding to different validity regimes: a shallow catalogue, fully constrained by existing observational datasets, and a deep catalogue extending the calibrated empirical model to better sensitivities, broadly consistent with future SKAO survey capabilities. 
We characterised the statistical properties of these catalogues, distinguishing between quantities directly constrained by the empirical tuning and emergent relations arising from the joint modelling of baryonic populations and dark-matter structure. 
The resulting mocks reproduce the targeted clustering and population statistics as expected.

We have thus obtained an empirically-driven description of the galaxy-halo connection, which only depends on the spatial distribution properties of the populations we are modelling.
The only assumption made is that, reasonably, galaxies are hosted in DM haloes and, implicitly, that the DM haloes in our simulated light-cone have physical properties compatible with those of the populations modelled.
In particular, we are making two implicit assumptions here:
\begin{enumerate}
    \item the mass resolution of our simulation is small enough to guarantee that our haloes can host all the sources in the populations we have modelled;
    \item the size of the original simulation box is large enough to have a fair representation of the clustering properties of the simulated galaxy populations.
\end{enumerate}
For the catalogues we have presented in this work, both of these assumptions are acceptable.
Nonetheless, the presented methodology can be applied to any catalogue of haloes and sub-haloes extracted from a simulated DM light-cone, thus opening to higher mass resolution and larger simulation boxes.

By construction, the simulations presented here do not rely on detailed physical prescriptions for galaxy formation, but instead encode current observational knowledge and its associated uncertainties. 
This makes them particularly well suited for survey forecasting and pipeline validation, as well as for exploring synergies between radio experiments and between radio and surveys at other wavelengths. 
The modular design of the pipeline allows individual components to be updated or replaced as new data and improved empirical constraints become available.

The framework naturally envisages a number of extensions. 
The mock catalogues can be generated for alternative cosmological models by exploiting the range of cosmologies available within the DEMNUni simulation suite or by adopting different dark-matter simulations. 
Additional baryonic components can be incorporated, including multi-frequency radio continuum emission through extended use of T-RECS or entirely new populations sampled from alternative empirical models. 
The inclusion of radio-selected counterparts at other wavelengths, for instance through spectral energy distribution modelling tools such as GalaPy \citep{Ronconi2024}, would enable more comprehensive multi-wavelength mocks. 
Finally, the availability of multiple populations on a common light-cone opens the way to cross-population forecasting studies and synergy analyses for future experiments, including joint applications with facilities such as ALMA, Euclid, and third-generation gravitational-wave detectors.
\section*{Data Availability}
All data presented in this work are available upon reasonable request.
We plan to upload the shallow and deep catalogues on Zenodo and the pipelines used to generate them on GitHub, along with the reviewed version of the manuscript.
The T-RECS library is publicly available on the GitHub page \href{github.com/abonaldi/TRECS}{github.com/abonaldi/TRECS}. The SCAMPy package is available on the GitHub page \href{github.com/TommasoRonconi/scampy}{github.com/TommasoRonconi/scampy}.
\begin{acknowledgements}
This work made use of the C++ \citep{stroustrup2013c++} and Python \citep{van2007python} programming languages, and of the following software: 
Astropy \citep{astropy2013}, NumPy \citep{harris2020numpy}, SciPy \citep{virtanen2020scipy}, Matplotlib \citep{Hunter2007matplotlib}, Emcee \citep{emcee2013}, pybind11 \citep{jakob2017pybind11}, HDF5 \citep{folk2011overview}, h5py \citep{collette2021h5py}.
\end{acknowledgements}
%
\bibliographystyle{aa} 
\bibliography{radiomocks} 

\begin{appendix}
\nolinenumbers



\section{Halo model predictions for the 1- and 2-point statistics}
\label{apx:halo_model}

Providing a detailed derivation of the halo model goes beyond the scope of this work, here we only list the statistics used in the main body of this manuscript and provide minimal derivation details. We recommend the interested reader to check on \cite{CooraySheth2002} and \cite{Asgari2023} for two good reviews.
The large scale statistics we tune our model against are the 1- and 2-point correlation functions.
In the halo-model formalism, the former is given by 
\begin{equation}
    \label{eq:1point_hm}
    n_g(z) \equiv \int_{M_\text{min}}^{M_\text{max}} \langle N_g \rangle (M_h)\; n(M_h, z)\; \dd{M_h}
\end{equation}
and, for a target population of galaxies at redshift $z$, it provides the average number density of objects hosted by DM haloes with masses, $M_h$, spanning the $M_\text{min}\div M_\text{max}$ range.

The 2 point statistics in the halo model are derived by Fourier transforming the non-linear power spectrum.
Given the first of the assumptions above, for a given redshift it is possible to separate the contributions to the 2-point statistics in the correlation between two objects belonging to the same DM halo (i.e. the 1-halo term) and the correlation between two objects belonging to two separate DM haloes (i.e. the 2-halo term): 
\begin{equation}
    \label{eq:pk_hm}
    P(k, z) = P_\text{1h}(k, z) + P_\text{2h}(k, z)\ \text{,}
\end{equation}
The first one of these terms reads
\begin{equation}
    \label{eq:pk1h_hm}
    \begin{split}
        P_\text{1h}(k, z) \equiv &\\
        = \dfrac{1}{n_g^2(z)}& \int_{M_{\text{min}}}^{M_\text{max}} \langle N_{\text{g}}(N_{\text{g}}-1)\rangle(M_h)\; n(M_h, z)\;  \bigl|\widetilde{u}(k, z| M_h)\bigr|^2 \dd{M_h}\text{,}
    \end{split}
\end{equation}
where $n(M_h, z)$ is a model for the DM halo mass function \citep[e.g.][]{SMT2001,ShethTormen2002,Despali2015,Lapi2020,Lapi2021,Lapi2022,Castro2023} while $\widetilde{u}(k, z| M_h)$ is the Fourier transform of the DM density profile \citep[e.g.][]{Burkert1995,NFW1997} and it, thus, depends on how matter is distributed within one halo and on how many haloes of a given mass are present within a volume in the Universe.

The 2-halo term on the other hand also depends on the linear matter power spectrum at the target redshift, $P_\text{lin}(k, z)$ and on a linear bias model \citep[e.g.][]{ShethTormen1999,SMT2001,Tinker2010,Castro2024}, $b(M_h, z)$:
\begin{equation}
    \label{eq:hm_pk2h_hm}
    \begin{split}
        P_\text{2h}(k, z) \equiv& \\
        = \dfrac{P_\text{lin}(k,z)}{n_g^2(z)} & \biggl[ \int_{M_{\text{min}}}^{M_\text{max}} \langle N_\text{g} \rangle(M_h)\; n(M_h)\; b(M_h, z )\;\widetilde{u}_h (k, z | M_h)\;\dd{M_h} \biggr]^2\text{.}
    \end{split} 
\end{equation}
This expression is an approximation not accounting for non-linearities.
Nevertheless, corrections to this approximation are mostly affecting the small-scales which are almost entirely dominated by the 1-halo component.

To get the real-space counterpart of the $P(k,z)$ we need to Fourier transform Eq.~\eqref{eq:pk_hm}:
\begin{equation}
    \label{eq:xi_hm}
    \xi(r, z) = \dfrac{1}{2\pi^2} \int_{k_\text{min}}^{k_\text{max}} \dd{k} k^2 P(k, z) \dfrac{\sin(k r)}{k r}\ \text{.}
\end{equation}
where $r$ is the separation between two sources. 

While Eq.~\eqref{eq:xi_hm} predicts a statistics on the 3-dimensional space, when working with galaxies it is not always possible to have a precise information on the third dimension (i.e. distance from the observer).
This is particularly true for radio-CoG for whom, due to the absence of prominent spectral features, redshift can be retrieved only from ancillary data that, on the other hand,  are typically not available for all the sources in the target radio survey.

Objects in a light-cone are therefore projected on a spherical surface, where the 1- and 2-point statistics result from the contribution of the objects up to the maximum depth reached by the survey.
For the 1-point statistics it is enough to integrate on redshift:
\begin{equation}
    \label{eq:ngtot_hm}
    n_g(z < z_\text{max}) = \int_0^{z_\text{max}} n_g(z) \dd z
\end{equation}
If we do not know the distance between us and the objects, we can measure transversal distances only in terms of angles, therefore we need to compute the \textit{angular} two-point correlation function (A2PCF) as
\begin{equation}
    \label{eq:hm_wt00}
        \omega ( \theta, z ) = \int \dfrac{\dd{V}(z)}{\dd{z}} \mathcal{N}^2(z)\; \omega[ r_p(\theta, z), z ] \dd{z}
\end{equation}
where $\dd{V}(z)\dd{z}$ is the co-moving volume unit, $\mathcal{N}(z)$ is the normalised redshift distribution of the target population, $r_p(\theta, z) = \theta\cdot d_C(z)$ with $d_C(z)$ the co-moving distance at redshift $z$ and $\omega[ r_p(\theta, z), z ]$ is obtained by means of a zeroth order Hankel transform of the power spectrum \citep{Hamilton2000}:
\begin{equation}
  \omega[ r_p(\theta, z), z ] = \mathcal{H}_0\bigl[P( k, z )\bigr]\text{.}
\end{equation}

\section{Dataset selection and SCAM model tuning}\label{apx:datasets}

In this Appendix we provide more details in the fitting performed to obtain the Halo Occupation Distribution (HOD) parameters reported in Tab.~\ref{tab:HOD}.

In order to obtain the simulated radio sky realization with injected clustering properties presented in the main body of this manuscript  we selected clustering measurements against which to fit the HOD parameters.
We have compared the selected measurements against the predictions of the halo model (see Section~\ref{sec:hod_model} and Appendix~\ref{apx:halo_model}).

We will list in this part the modelling choices common to both the shallow and deep catalogue.
In the next two sub-sections we will instead detail the choices which are specific of the two different simulations flavours presented.

\begin{enumerate}
    \item the \textbf{HOD parameters} to fit are the same for both the populations of sources present in the catalogues, as well as the formula for the average number of central galaxies hosted in each halo of given mass $M_h$, which is given by 
    \begin{align}
    \label{eq:Ncen_CoG+HIG}
    N_\text{cen}(M_h) &= \dfrac{1}{2}\left[1 + \erf\left(\dfrac{\log M - \log M_{\text{min}}}{\sigma}\right)\right]\,\text{.}
    \end{align}
    The average number of satellite galaxies is instead modelled differently in the two populations, with
    \begin{enumerate}
        \item \textbf{Continuum galaxies:} 
        \begin{align}
            \label{eq:Nsat_CoG}
            N_\text{sat}(M_h) &= N_\text{cen}(M_h)\cdot\left(\dfrac{M_h - M_\text{0}}{M_1}\right)^{\alpha}
        \end{align}
        \item \textbf{HIG:}
        \begin{align}
            \label{eq:Nsat_HIG}
            N_\text{sat}(M_h) &= \left(\dfrac{M_h - M_\text{0}}{M_1}\right)^{\alpha}
        \end{align}
        where we do not multiply for Eq.~\eqref{eq:Ncen_CoG+HIG} in order to account for the possibility of having HIG in a haloes where no central HIG is present. 
        This is a reasonable assumption, as central galaxies are typically more massive and typically elliptical, with low content of \HI.
    \end{enumerate}
    \item All our modelling is performed assuming a \textbf{flat $\boldsymbol{\Lambda}$CDM cosmology} with approximated \cite{Planck2013} parameters, in order to match the same cosmological parameters used to run the DEMNUni simulations. 
    This is a consistency assumption; should we change background cosmological simulation, we should re-run the fitting by changing the cosmological parameters accordingly.
    Nevertheless, it has been shown \citep[e.g.][]{Contreras2023} that the HOD parameters depend mildly on the background cosmology and more on the astrophysical properties of the population modelled \citep[e.g.][]{Perez2024}.
    \item We fit continuum and line-emitting galaxies separately and independently on different datasets.
    We then also populate the haloes without informing one species about the other.
\end{enumerate}

What differentiates the two catalogues is, \textit{(i)} the dataset against which the model is fitted, \textit{(ii)}, the prior volume explored and, \textit{(iii)}, the likelihood used to explore the posterior space.
\begin{table}
    \centering
    \begin{tabular}{ccccc}
\toprule\midrule
$\log M_\text{min}$ & $\log \sigma$ & $\log M_\text{0}$ & $\log M_\text{1}$ & $\log \alpha$\\
\midrule
$[10.4, 16]$ & $[-3, +2]$ & $[10.4, 16]$ & $[10.4, 16]$ & $[-3, +2]$\\
\bottomrule\bottomrule
\end{tabular}
    \caption{Flat uniform prior limits for the HOD parameters used in all the parameter space samplings presented in this work.}
    \label{tab:HODprior}
\end{table}

\subsection{Shallow catalogue}\label{apx:shallow_fit}

\begin{description}
    \item \textbf{Dataset:} for CoG, plenty of measurements of the angular 2 point correlation function (A2PCF) can be retrieved from literature \citep[][]{Magliocchetti2017,Hale2018,RanaBagla2019,Chakraborty2020,Bonato2021,Mazumder2022,Hale2024}. We fit our model on the A2PCF of all sources in the catalogue provided by \cite{Hale2018} as they provide measurements in the widely observed COSMOS field. Furthermore, they also divide the dataset and measurement divided between SFG and AGN, allowing us for cross-checking the resulting clustering properties, during the preparation of this work.
    Measurements of the two-point correlation function (2PCF) of HIG are more scarcely available with respect to their continuum equivalent.
    We could only retrieve data measured on the ALFALFA survey \citep{Martin2012,Papastergis2013}. We use the three-dimensional measurement of \cite{Martin2012}.
    
    \item \textbf{Prior:} The prior for both continuum and HIG has a top-hat shape (flat uniform probability), with intentionally wide boundaries and is only limited by the background dark matter simulation. We report the prior limits in Tab.~\ref{tab:HODprior}, all parameter space dimensions have been explored in log-space.
    In the shallow case, the only information provided to this prior is in the characteristic masses ($M_\text{min}$, $M_0$ and $M_1$) which are limited to vary between the minimum ($\min{\left[\log M_\text{halo}\right]} \sim 10.4$) and maximum ($\max{\left[\log M_\text{halo}\right]} \sim 16$) halo mass of the original DM simulation.
    \item \textbf{Likelihood:} Since it is not possible to obtain a precise redshift estimate for CoG, we fit the model against the projected A2PCF, while the information on the redshift distribution is included as a separate term in the Gaussian likelihood:
    \begin{equation}
        \label{eq:logL_CoG_shallow}
        \log \mathcal{L}_\text{CoG}(\boldsymbol{p}) = \dfrac{1}{2}\left[\chi^2_{w(\theta)} + \chi^2_{\mathcal{N}(z)} + \chi^2_{n_g^\text{tot}}\right]\,\text{,}
    \end{equation}
    where the first term in the right-hand side of the equation is the chi-squared of the A2PCF, the second term the redshift distribution and the third term the total surface density.
    For HIG we instead have access to the 2PCF in three-dimension and the Gaussian log-likelihood is instead:
    \begin{equation}
        \label{eq:logL_HIG_shallow}
        \log \mathcal{L}_\text{HIG}(\boldsymbol{p}) = \dfrac{1}{2}\left[\chi^2_{\xi(r)} + \chi^2_{n_g(z)}\right]\,\text{,}
    \end{equation}
\end{description}

\subsection{Deep catalogue}\label{apx:deep_fit}

\begin{figure*}
    \centering
    \includegraphics[width=0.98\linewidth]{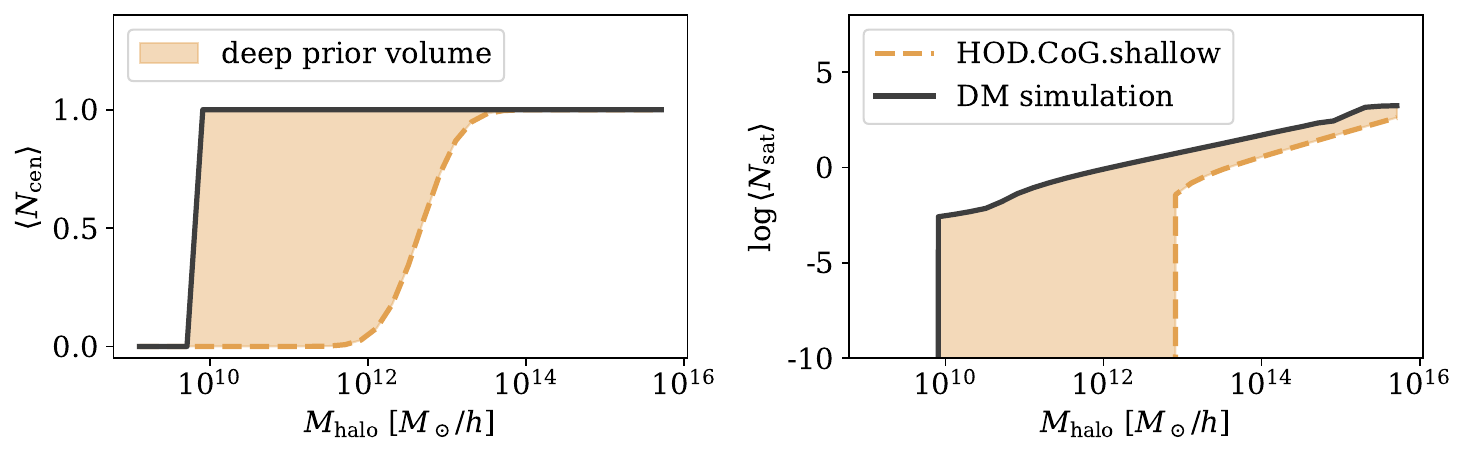}\\
    \includegraphics[width=0.98\linewidth]{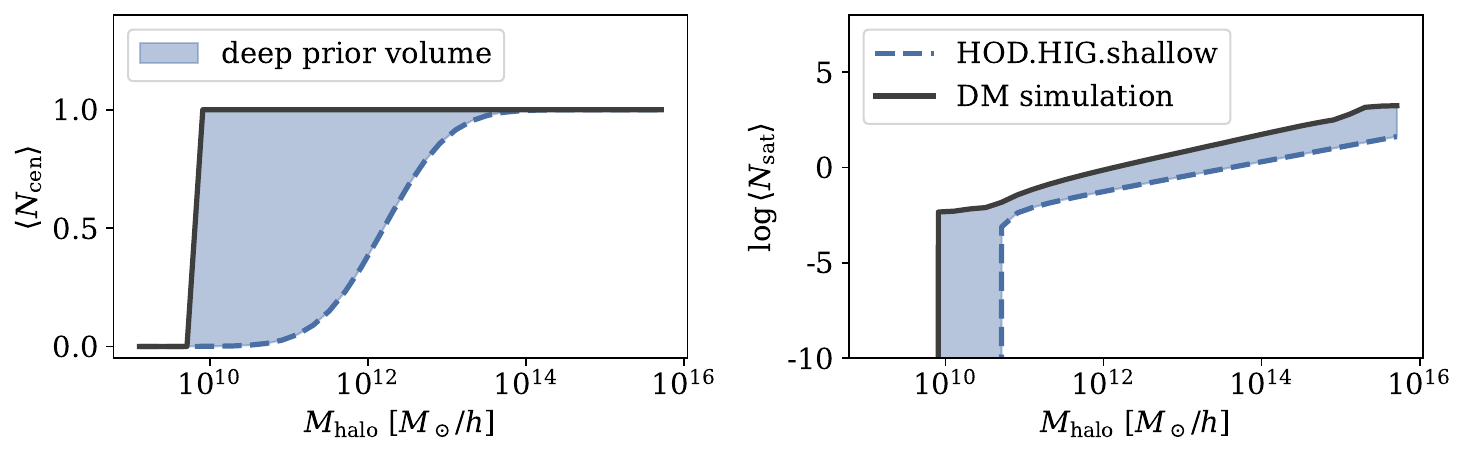}
    \caption{Priors (shaded coloured regions) on the occupation probabilities, $N_\text{cen}(M_h)$ and $N_\text{sat}(M_h)$, imposed to the deep catalogue HOD parameters fitting. The upper panels mark the prior on CoG (results shown in Fig.~\ref{fig:CoG_NcNs}) while the lower panels show the prior on \HI sources (results shown in Fig.~\ref{fig:HIG_NcNs}). The upper limit (black solid line) corresponds to the average abundance of centrals and satellites measured in the DM simulation while the lower limit (coloured dashed lines) are given by the best fit HOD model obtained for the shallow catalogue.}
    \label{fig:priors_NcNs}    
\end{figure*}
\begin{description}
    \item \textbf{Dataset:} the deep catalogue is tuned against an extrapolation to meet the performances forecasted for future SKAO surveys. We have built a catalogue of mock sources using the T-RECS tool \citep{Bonaldi2018,Bonaldi2023} using the flux limits reported in \cite{CosmoSWG2020} for expected continuum surveys and from \cite{Hartley2023} for expected HIG surveys. We have then extracted the redshift distribution of both the populations of objects and used those as mock-datasets.
    \item \textbf{Prior:} The prior on the parameters used for the shallow catalogue has the same functional form (i.e. a top-hat) and boundaries (see Tab.~\ref{tab:HODprior}) used for the shallow catalogue. In the deep catalogue though we make a further assumption: besides imposing an upper limit not to over-shoot the average number of available sub-haloes (centrals and satellites) in the background DM simulation (black solid lines in Fig.~\ref{fig:priors_NcNs}), we also make the assumption that, on average, all the haloes which can host an object in the shallow catalogue (dashed coloured lines in Fig.~\ref{fig:priors_NcNs}), can also host an object in the deep catalogue.
    In order to inform our MCMC sampling about this choice, we include in the prior computing function the condition to have occupation probabilities belonging to the shaded regions shown in Fig.~\ref{fig:priors_NcNs}.
    \item \textbf{Likelihood:} since no correlation data are available in this case, we have simplified the likelihoods to only depend on the 1-point simulated data.
    We thus use the Gaussian likelihood:
    \begin{equation}
        \label{eq:logL_CoG_deep}
        \log \mathcal{L}_\text{CoG}(\boldsymbol{p}) = \dfrac{1}{2}\left[\chi^2_{\mathcal{N}(z)} + \chi^2_{n_g^\text{tot}}\right]\,\text{,}
    \end{equation}
    for the continuum case and 
    \begin{equation}
        \label{eq:logL_HIG_deep}
        \log \mathcal{L}_\text{HIG}(\boldsymbol{p}) = \dfrac{1}{2}\chi^2_{n_g(z)}\,\text{,}
    \end{equation}
    for the HIG case.
\end{description}

\section{Result of the SHAM step}\label{apx:sham}

\begin{figure*}
    \centering
    \includegraphics[width=0.98\textwidth]{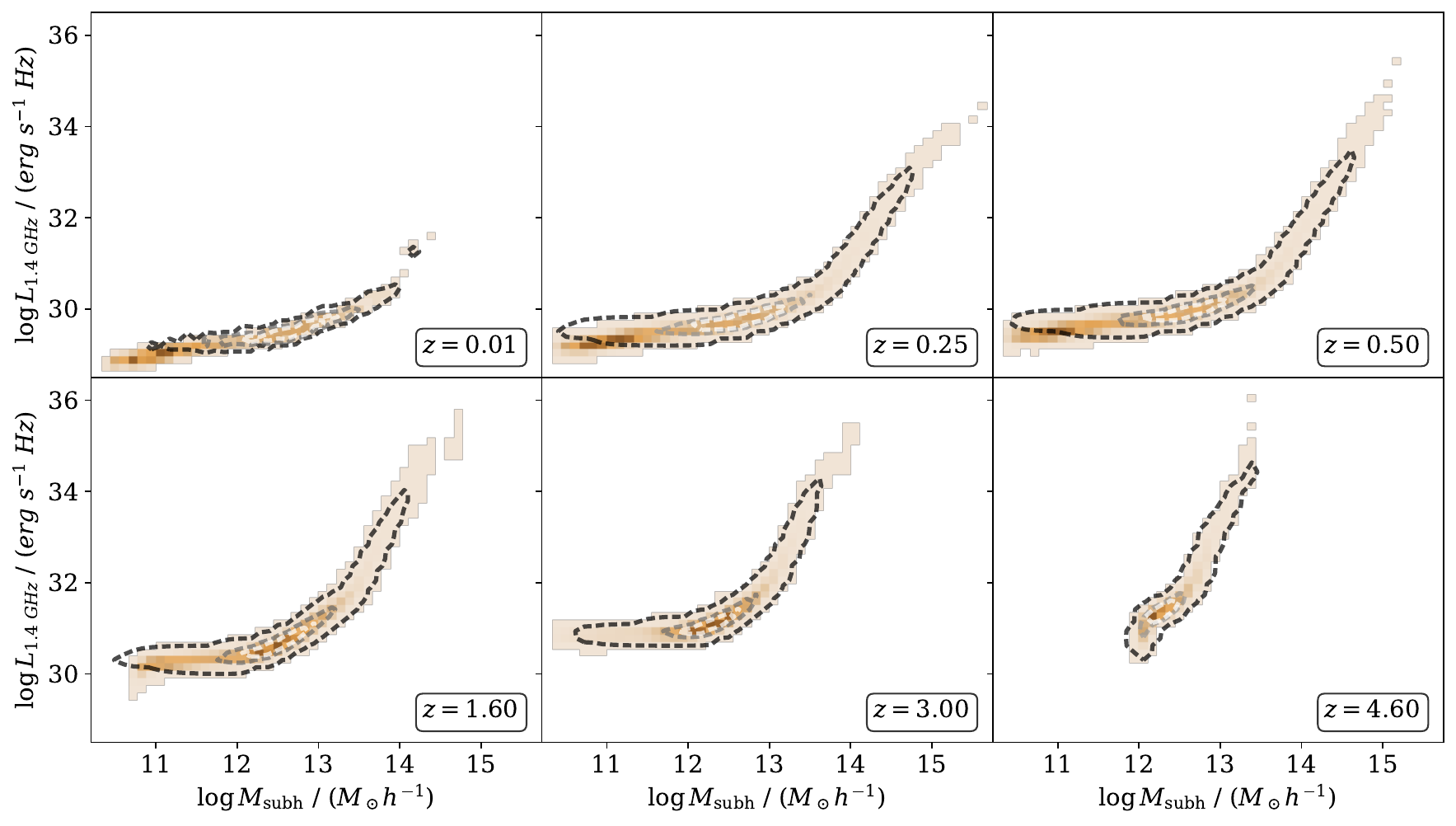}
    \caption{Result of the SHAM algorithm for the CoG at varying redshift. We have chosen a sub-sample of the 41 redshift slices available from $z_\text{min} = 0.01$ to $z_\text{max} = 5.0$. The coloured 2D histogram shows the distribution in the deep catalogue while the gray contours the distribution in the shallow catalogue.}
    \label{fig:CoG_SHAM_ss}
\end{figure*}
\begin{figure*}
    \centering
    \includegraphics[width=0.98\textwidth]{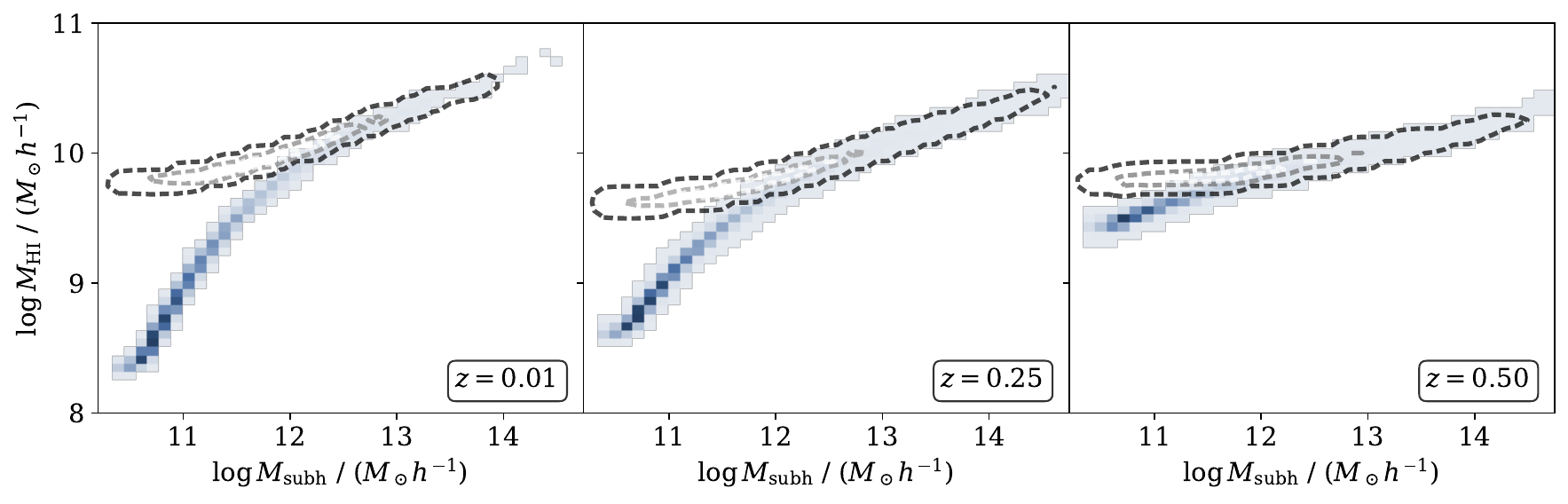}
    \caption{Same as Fig.~\ref{fig:CoG_SHAM_ss} for the HIG catalogue. In this case we only show the first 3 panels as HIG are populated only up to redshift $z_\text{max}=0.5$.}
    \label{fig:HIG_SHAM_ss}
\end{figure*}
In Fig.~\ref{fig:CoG_SHAM_ss} and Fig.~\ref{fig:HIG_SHAM_ss} we show the result of the second step of the SCAM algorithm: matching the sub-haloes selected with the CoG from T-RECS.
The different panels show the result of scattering the luminosity of CoG at $1.4$ GHz (Fig.~\ref{fig:CoG_SHAM_ss}) and the \HI mass (Fig.~\ref{fig:HIG_SHAM_ss}) against the mass of the underlying DM sub-halo, in different shells centred at increasing values of redshift (as reported in the text-box on the lower right side of each panel).
The coloured regions in the panels show the distribution of sources in the deep catalogue while the grey contours mark the distribution in the shallow catalogue. 
As expected from Fig.~\ref{fig:CoG_NcNs}, the two distributions do not differ much.
Most of the differences between the two distributions and thus, between the deep and shallow catalogues, are concentrated at the lowest redshift values.
This is not surprising, as cosmological N-body simulations tend to predict more satellite sub-haloes at lower redshift, and our HOD model predicts that the 2 catalogues differ only in the satellite galaxy number.

Going deeper in hydrogen is also reflected into the substantial difference shown in Fig.~\ref{fig:HIG_SHAM_ss}. 
In the Figure, the blue shades mark the scatter between \HI mass content within galaxies against their underlying DM budget. 
At lower redshift, where the largest number of galaxies are sampled, the difference with the grey dashed contours marking the distribution in the shallow catalogue, is larger and diminishes in magnitude as the simulation moves to larger redshift values (reported in the lower right box of each panel).
\end{appendix}
\end{document}